\documentclass[traditabstract]{aa} % for the abstract without structuration 
\usepackage{graphicx}
\usepackage{txfonts}
\begin{document}
   \title{The old and heavy bulge of M31}

   \subtitle{I. Kinematics and stellar populations}

   \author{R.P. Saglia\inst{1,2}
           \and
           M. Fabricius\inst{1,2}
           \and
           R. Bender\inst{1,2}
           \and
           M. Montalto\inst{2,1}
           \and
           C.-H. Lee\inst{2,1}
           \and
           A. Riffeser\inst{2,1}
          \and
          S. Seitz\inst{2,1}
          \and
          L. Morganti\inst{1,2}
          \and
          O.~Gerhard\inst{1,2}
          \and
          U. Hopp\inst{2,1}
        }

\institute{Max-Planck Institut f\"ur extraterrestrische Physik, 
           Giessenbachstrasse, Postfach 1312, D-85741 Garching, Germany\\
           \email{saglia@mpe.mpg.de}
           \and
           Universit\"ats-Sternwarte M\"unchen, Scheinerstrasse 1, D-81679 
           M\"unchen, Germany \\
           }

   \date{Received ; accepted }

% \abstract{}{}{}{}{} 
% 5 {} token are mandatory
 
   \abstract{We present new optical long-slit data along 6 position
     angles of the bulge region of M31. We derive accurate stellar and
     gas kinematics reaching 5 arcmin from the center, where the disk
     light contribution is always less than 30\%, and out to 8 arcmin
     along the major axis, where the disk makes 55\% of the total
     light. We show that the velocity dispersions of McElroy (1983)
     are severely underestimated (by up to 50 km/s). As a consequence,
     previous dynamical models have underestimated the stellar mass of
     M31's bulge by a factor 2. As a further consequence, the
     light-weighted velocity dispersion of the galaxy grows to 166
     km/s and to 170 km/s if also rotation is taken into account, thus
     reducing the discrepancy between the predicted and measured mass
     of the black hole at the center of M31 from a factor 3 to a
     factor 2. The kinematic position angle varies with distance,
     pointing to triaxiality, but a quantitative conclusion can be
     reached only after simultaneous proper dynamical modeling of the
     bulge and disk components is performed. We detect gas
     counterrotation near the bulge minor axis. We measure eight
     emission-corrected Lick indices. They are approximately constant
     on circles.  Using simple stellar population models we derive the
     age, metallicity and $\alpha$-element overabundance profiles.
     Except for the region in the inner arcsecs of the galaxy, the
     bulge of M31 is uniformly old ($\ge 12$ Gyr, with many best-fit
     ages at the model grid limit of 15 Gyr), slightly
     $\alpha$-elements overabundant ($[\alpha/Fe]\approx 0.2$) and at
     solar metallicity, in agreement with studies of the resolved
     stellar components.  The predicted u-g, g-r and r-i Sloan color
     profiles match reasonably well the dust-corrected observations,
     within the known limitations of current simple stellar population
     models.  The stellar populations have approximately radially
     constant mass-to-light ratios ($M/L_R\approx
     4-4.5M_\odot/L_\odot$ for a Kroupa IMF), in agreement with
     stellar dynamical estimates based on our new velocity
     dispersions. In the inner arcsecs the luminosity-weighted age
     drops to 4-8 Gyr, while the metallicity increases to above 3
     times the solar value. Starting from 6 arcmin from the center
     along the major axis, the mean age drops to $\le 8$ Gyr, with
     slight supersolar metallicity ($\approx +0.1$ dex) and
     $\alpha-$element overabundance ($\approx +0.2$ dex), for a
     mass-to-light ratio $M/L_R\le 3M_\odot/L_\odot$.  Diagnostic
     diagrams based on the [OIII]/H$\beta$ and [NI]/H$\beta$ emission
     line equivalent widths (EWs) ratios indicate that the gas is
     ionized by shocks outside 10 arcsec, but an AGN-like ionizing
     source could be present near the center. We speculate that a
     gas-rich minor merger happened some 100 Myr ago, causing the
     observed minor axis gas counterrotation, the recent star
     formation event, and possibly some nuclear activity.}
  % context heading (optional)
  % {} leave it empty if necessary  
  % {}
  % aims heading (mandatory)
  % {}
  % methods heading (mandatory)
  % {}
  % results heading (mandatory)
  % {}
  % conclusions heading (optional), leave it empty if necessary 
  % {}

   \keywords{Galaxies: individual: M31, kinematics and dynamics, 
stellar content}

   \maketitle
%
%________________________________________________________________

\section{Introduction}
\label{sec_intro}
This is the first of two papers presenting new optical spectra for the
bulge of M31 to study its stellar populations and assess its
triaxiality through dynamical modeling. Here we present the new data and
constrain the stellar populations.

In the past 50 years papers studying the dynamics of our neighbour
galaxy M31 have been published on a regular basis, discussing gas
kinematics, both by optical spectroscopy (Boulesteix et al.
\cite{Boulesteix87}, Pellet \cite{Pellet76}), and in HI (Kent
\cite{Kent89a}, Braun \cite{Braun91}, Chemin et al. \cite{Chemin09}
and references therein), stellar kinematics concentrating on the
central regions to probe the black hole dynamics (Bender et al.
\cite{Bender05}) or considering the whole bulge (McElroy
\cite{McElroy83}). The data are used to construct dynamical models of
the galaxy (Widrow et al. \cite{Widrow03}, Klypin, Zhao and Somerville
\cite{Klypin02}) and possibly probe the tridimensional distribution of
its stellar components.  The question of the triaxiality of M31 bulge
has been posed early on (Stark \cite{Stark77}, Gerhard
\cite{Gerhard86}) and is of significant importance for the
understanding of M31, but a definitive quantitative modeling of both
photometry and kinematics is still missing. A bar could also be
present (Athanassoula \& Beaton \cite{Athanassoula06}, Beaton et al.
\cite{Beaton07}).  Moreover, investigations of the stellar populations
of the central regions of M31 through the measurement of Lick indices
have been performed (Davidge \cite{Davidge97}). They indicate the
presence of a young and metal rich population in the inner arcsecs of
the galaxy.  Parallely, studies of the resolved stellar population of
the bulge of M31 have assessed that the global stellar population of
the M31 bulge must be as old as the bulge of Milky Way, resolving
previous claims of younger ages as due to crowding problems (Stephens
et al.  \cite{Stephens03}).

Two considerations convinced us of the necessity to collect new
optical spectroscopic information for the bulge of M31, in addition to
the old age of the dataset of McElroy (\cite{McElroy83}). The first
one is the start of PAndromeda, an extensive monitoring campaign of
M31 with the PanSTARRS-1 telescope and camera system (Kaiser
\cite{Kaiser04}), that is in principle able to deliver hundreds of
pixel lensing events, probing both bulge and disk regions. Detailed
stellar population and dynamical models, based on accurate spectral
information, are needed to interpret these events as due to a compact
baryonic dark matter component (the so-called MACHOs) rather than
self-lensing of stellar populations (Kerins et al. \cite{Kerins01},
Riffeser et al.  \cite{Riffeser06}).  The second is the development of
new modeling techniques of both simple stellar populations and stellar
dynamical systems. On the one hand, the interpretation of Lick indices
(Worthey et al \cite{Worthey94}) in terms of the most recent simple
stellar population models (Maraston \cite{Maraston98},
\cite{Maraston05}) that take into account the variation of
$[\alpha/$Fe] (Thomas, Maraston and Bender \cite{TMB03}), allows the
accurate determination of the stellar population ages, metallicities
and overabundances, and therefore the prediction of stellar
mass-to-light ratios. On the other hand, new dynamical modeling codes,
like N-MAGIC (De Lorenzi et al.  \cite{DeLorenzi07}) allow the
flexible dynamical modeling of triaxial structures, optimally
exploiting the information contained in the line-of-sight velocity
distributions that modern programs for the analysis of the galaxy
optical spectra are able to extract (Bender, Saglia and Gerhard
\cite{BSG94}), well beyond the mean velocities and velocity
dispersions of McElroy (\cite{McElroy83}).

In the following we discuss our new spectroscopic observations of the
bulge of M31. A future paper (Morganti et al.
\cite{Morganti09}) will report on the dynamical modeling. In Sect.
\ref{sec_obs} we present the observations and the data reduction.  In
Sec. \ref{sec_kinematics} we derive the stellar and gas kinematics and
the strengths of the absorption and emission lines. In Sec.
\ref{sec_mod} we discuss the modeling of the new spectroscopic data.
We analyze the stellar population in Sect. \ref{sec_stelpop} and
discuss previous axisymmetric dynamical models of the bulge of M31 in
Sect.  \ref{sec_dyn}. Sect. \ref{sec_ionize} considers the possible
excitation sources compatible with the observed emission line EW
ratios. We draw our conclusions in Sect.  \ref{sec_conc}.

\section{Observations and Data reduction}
\label{sec_obs}

\begin{figure}%[h!]
\centering
\includegraphics[width=9cm]{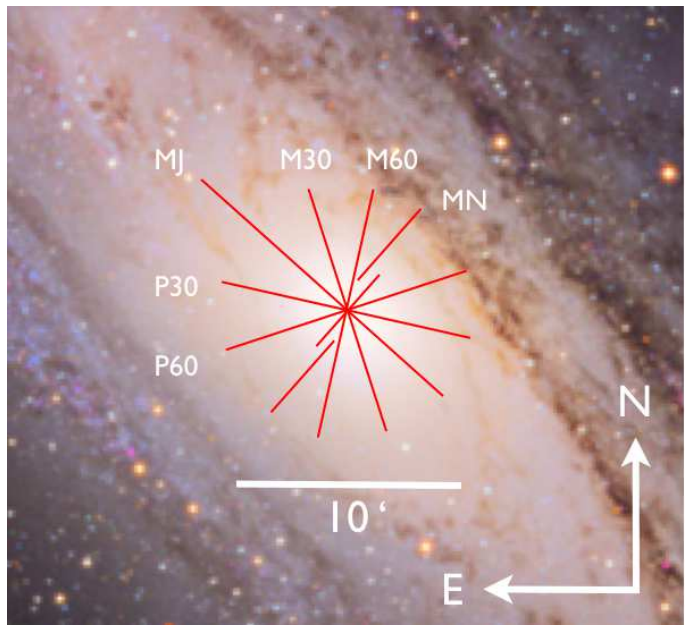}
\caption{The positions of the 18 M31 slit pointings superimposed on a
  NOAO image of the Andromeda galaxy (credit Adma
  Block/NOAO/AURA/NSF). See Table \ref{tab_log} for the naming
  convention.}
\label{fig_slits}
\end{figure}

We observed the bulge of M31 using the Low Resolution Spectrograph
(LRS, Hill et al. \cite{Hill98}) at the Hobby-Eberly Telescope
equipped with a 1.5 arcsec wide, 3.5 arcmin long slit, the E2 grism
and a Ford Aerospace CCD device, with 3072$\times$1024 15$\mu$m pixels
(usable range 2750$\times$900 pixels) and a scale of 0.235 arcsec per
pixel.  We covered the wavelength range $\lambda=4787-5857$ \AA\ with
0.36 \AA\ per pixel and an instrumental resolution of
$\sigma_{inst}=57$ km/s.  During the period 10th-19th August 2007 (see
Table \ref{tab_log}) M31 was observed in service mode along 6 position
angles. The seeing varies from 1.3 to 2.5 arcsec. At each position
angle three 10 minutes exposures were taken, the first one centered on
the galaxy, the second one shifted to the west (decreasing RA) 3.5
arcmin in the slit direction, and the third one shifted to the east
(increasing RA) 3.5 arcmin in the slit direction. Finally, on the 15th
of September 2009 we collected a 20 minutes exposure along the major
axis, shifted 5 arcmin from the center in the eastern direction. Fig.
\ref{fig_slits} shows the distribution of the 19 slit M31 pointings on
the sky with the naming convention given in Table \ref{tab_log}, where
MJ shortens for the major axis and MN for the minor axis. The position
angle of the bulge MJ axis $PA=48$ is fixed from our 2MASS photometry
(see Sect. \ref{sec_dyn} and Fig. \ref{fig_triaxial}). Note that this
is different from the position angle of the major axis of the disk of
M31 ($PA=38$, de Vaucoulers \cite{devauc58}). Apart from the MN
pointings, the slits with a given position angle are always well
aligned and with small overlaps, providing spectra out to 5 arcmin
from the center. In addition, the MJEE slit probes the major axis out
to 8 arcmin from the center. The MNE and MNW are slighly shifted from
the MNC central slit (by 34'' orthogonal to the slit to the west and
to the east respectively).  Furthermore, 10 minutes exposures of empty
sky regions were also taken, as well as several kinematic and Lick
standard stars, wiggled and trailed along the slit.

\begin{table}
\caption{Log of the observations.}             % title of Table
\label{tab_log}      % is used to refer this table in the text
\centering                          % used for centering table
\begin{tabular}{c l r c l}        % centered columns (4 columns)
\hline\hline                 % inserts double horizontal lines
Date      & Name   & PA  & FWHM & Comment \\    % table heading 
          &      & (deg.) &  (``) & \\
\hline                                    % inserts single horizontal line
20070812 & MJC    &  48  & 1.40 & Centered major axis \\ 	
20070812 & MJE    &  48  & 1.46 & Eastern  major axis \\	 
20070812 & MJW    &  48  & 1.38 & Western  major axis \\	
20090915 & MJEE   &  48  & 2.13 & Eastern  major axis \\	 
20090915 & SKYMJEE &  48 & -    & Empty sky\\	 
20070813 & P30C   &  78  & 1.59 & Centered $+30^\circ$ \\	 
20070813 & P30E   &  78  & 1.91 & Eastern  $+30^\circ$ \\	 
20070813 & P30W   &  78  & 2.08 & Western  $+30^\circ$ \\	 
20070813 & P30SKY &  90  & -    & Empty sky \\	 
20070814 & MNC    & 138  & 1.28 & Centered minor axis \\ 	
20070814 & MNE    & 138  & 1.36 & Eastern  minor axis, shifted \\	 
20070814 & MNW    & 138  & 1.41 & Western  minor axis, shifted \\	
20070814 & MNSKY  &  90  & -    & Empty sky \\  
20070816 & P60C   & 108  & 1.61 & Centered $+60^\circ$ \\	 
20070816 & P60E   & 108  & 1.57 & Eastern  $+60^\circ$ \\	 
20070816 & P60W   & 108  & 1.63 & Western  $+60^\circ$ \\	 
20070816 & P60SKY &  90  & -    & Empty sky \\	 
20070817 & M30C   &  18  & 2.53 & Centered $-30^\circ$ \\	 
20070817 & M30E   &  18  & 2.21 & Eastern  $-30^\circ$ \\	 
20070817 & M30W   &  18  & 2.21 & Western  $-30^\circ$ \\	 
20070819 & M60C   & 168  & 1.43 & Centered $-60^\circ$ \\	 
20070819 & M60E   & 168  & 1.78 & Eastern  $-60^\circ$ \\	 
20070819 & M60W   & 168  & 1.94 & Western  $-60^\circ$ \\	 
20070819 & M60SKY &  90  & -    & Empty sky \\	 
\hline                                   %inserts single line
\end{tabular}
\end{table}

\begin{figure*}%[t!]
\centering
\includegraphics[width=18cm]{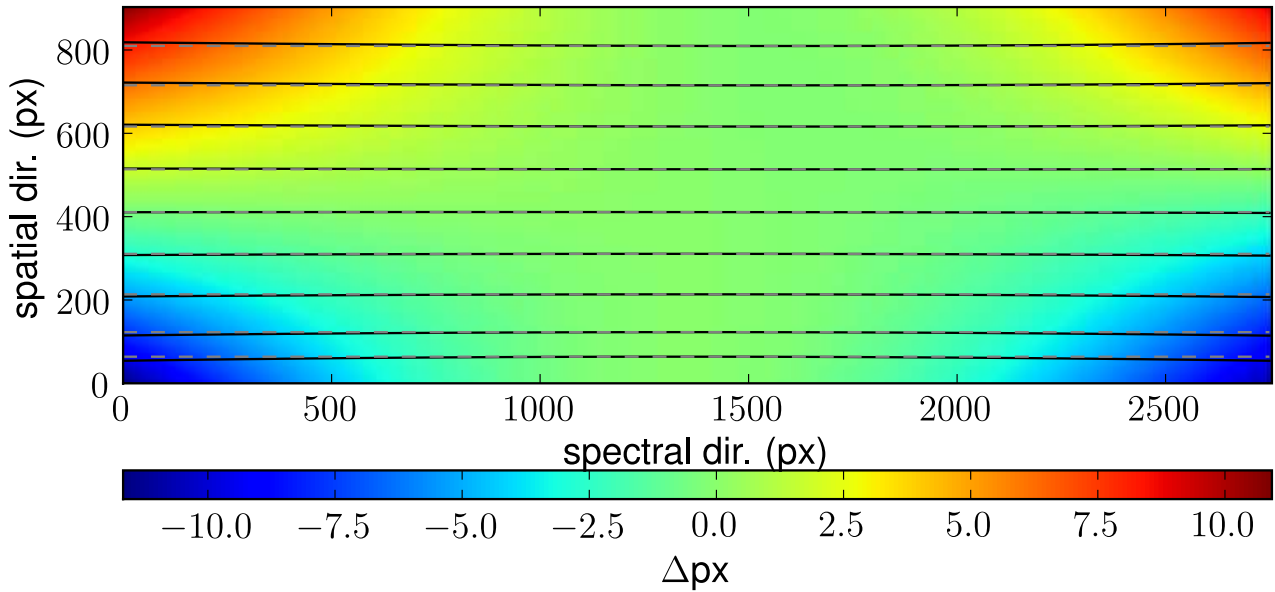}
\caption{The anamorphic distortion of the LRS as mapped by tracing the
  peaks of the stellar spectra. The traces of 9 stellar spectra are
  shown as full black lines and are compared to the corresponding
  straight lines (dashed). The interpolated differences (in pixel)
  between the stellar traces and the Cartesian grid as a function of
  the position on the chip are shown coded by colour. }
\label{fig_anamorph}
\end{figure*}

The data reduction followed the usual procedure described in Mehlert
et al. (\cite{Mehlert00}) under MIDAS with some additional steps
needed to correct for the spectral alignment and the anamorphism of
the LSR spectrograph. After bias subtraction and flatfielding, the
2$^\circ$ tilt between the wavelength direction and the CCD was
removed through appropriate rebinning. Two bad columns at positions
corresponding to $\lambda\approx 4850$\AA\ were corrected by
interpolation. The wavelength calibration used Cd and Ne calibration
lamps frames taken at the end of each night, and a 3rd order
polynomial fitting 9 lines and giving an rms of less than one pixel.
Residual overall shifts in the starting wavelength of the wavelength
calibrated science frames were corrected by referencing to the 5577
\AA\ sky line. After filtering cosmic ray hits, the science frames
were rebinned to a logarithmic wavelength scale. The sky subtraction
for the M31 frames was performed through the four empty sky frames
available.  The averaged sky continuum amounted to 3.5 counts per
pixels (varying by at most 10\% from frame to frame). This is only
14\% of the averaged flux measured at the extreme ends of the MJE and
MJW slits, 23\% at the end of the MJEE slit, but 44\% at those of the
MNE and MNW slits.  In contrast, the flux per pixel in the strongest
sky line at 5577 \AA\ is 160 counts with variations to up to 50\% from
frame to frame. We constructed a line-free and a line-only average sky
spectrum for each of the available sky frames and subtracted it with
proper scaling from the corresponding science frames. The average of
the available sky frames was used for the nights without empty sky
observations. Stars were extracted tracing the peaks corresponding to
their trails across the slit as a function of wavelength and averaging
over a 10 pixel wide window. The final spectra are averages of the
extracted ones.  This procedure also allowed us to map the anamorphic
distortions of the spectrograph, that at the short and long
wavelengths bend the spectra upwards in the upper end and downwards in
the lower end of the slit (see Fig.  \ref{fig_anamorph}). Moreover,
the sky was estimated at the ends of the slit and subtracted. Finally,
the anamorphic mapping was used to rebin the sky subtracted frames of
M31 on a cartesian grid.  Finally, the M31 spectra were rebinned in
the radial direction to give approximately constant signal-to-noise
ratio.

\section{Kinematics and Line Strengths}
\label{sec_kinematics}

The stellar kinematics were derived using the Fourier Correlation
Quotient technique (Bender \cite{Bender90}), with the addition of an
iterative procedure to correct for the emission clearly detected in
the H$\beta$ line and the [OIII] and [NI] doublets, similar to the
method of Sarzi et al. (\cite{Sarzi06}). The best fitting stellar
template convolved to the derived line-of-sight velocity distribution
(LOSVD) is subtracted from the galaxy spectra. Five Gaussian functions
are fitted to the residuum spectrum, in correspondence to the two
doublets and the H$\beta$ line.  Six parameters are derived: the
recessional velocity of the gas emission and its velocity dispersion
(determined by fitting just the [OIII] lines), plus 5 normalization
factors. The best-fitting Gaussians are subtracted from the original
spectrum and the stellar kinematics is derived a second time, using
the wavelength range $4870-5470$~\AA . The final parameters of the gas
emission are derived by fitting the original spectrum where the
improved, iterated best-fitting stellar kinematic template is
subtracted. The best-fitting template is chosen from a pool of single
star spectra and simple stellar population models from Vazdekis
(\cite{Vazdekis99}), by minimizing the $\chi^{2}$ between the galaxy
and the LOSVD-convolved template, and results in an almost
mismatch-free description of the galaxy spectrum. An 8th order
polynomial is used to subtract the continuum and the first and last
three channels in Fourier space are filtered out. Monte Carlo
simulations as described in Mehlert et al.  (\cite{Mehlert00}) are
performed to derive the final errors on the estimated parameters and
estimate the size of systematic errors. With the choice of continuum
fitting and filtering given above the systematic errors are always
smaller than the statistical ones.  We have tested that none of the
results presented below depend on the precise order of the continuum
subtraction within the errors, as soon as it is larger than 6. The
measured kinematics are very precise, with statistical errors on
velocities of the order of 2 km/s, of the order of 3 km/s (i.e. 3\%),
on $\sigma$, less than 0.02 on $H_3$ and $H_4$. We investigated the
systematic effects of sky subtraction by deriving the kinematics from
frames where $\pm10$\% of the sky was additionally subtracted. At the
extreme ends of the slits differences in velocities up to 5 km/s, in
$\sigma$ up to 12 km/s, in $H_3$ up to 0.03 and in $H_4$ up to 0.05
are seen.

Fig. \ref{fig_kin} shows the derived stellar line of sight recessional
velocity $V$, velocity dispersion $\sigma$ and Hermite-Gauss
coefficients $H_3$ and $H_4$ following Bender, Saglia and Gerhard
(\cite{BSG94}). Overall, the kinematic data are symmetric with respect
to the center within the errors. Stellar rotation is detected at some
level at all position angles. Along the major axis (MJ) it keeps
rising with increasing distance from the center, reaching $\approx
100$ km/s at the last measured point at $\approx 8$ arcmin, where the
disk light dominates.  The velocity dispersion reaches a maximum of
170 km/s at $\approx 60$ arcsec from the center. Along or near the
major axis, where stellar rotation is clearly detected, the $H_3$
parameter anticorrelates (as expected, see Bender et al.
\cite{BSG94}) with the stellar velocity out to $\approx 100$ arcsec,
to change sign at larger distances (see discussion in Sect.
\ref{sec_dyn}). In the inner arcsecs the signature of the central
supermassive black hole is clearly visible, with a steep increase in
velocity and velocity dispersion. In the inner region, the comparison
with the kinematics of van der Marel et al.  (\cite{vdMarel94}) along
the major and minor axis shows overall good agreement; however they
measure systematically ($\approx0.05$) lower values of $H_4$. Since we
agree with the data of Kormendy and Bender (\cite{Kormendy99}) and we
do not expect residual systematics from our Monte Carlo tests (see
above), we trust our $H_4$ profiles as the correct ones.  The
comparison with McElroy (\cite{McElroy83}) shows that this old dataset
gives seriously underestimated velocity dispersions. The smallest
differences are observed along the minor axis. The discrepancy is
probably due to the sensitivity of McElroy analysis's method to the
disk light contamination (see discussion in Sect. \ref{sec_dyn}).  The
recessional velocities are in reasonable agreement.

\begin{figure*}%[t!]
\centering
\includegraphics[width=14cm,angle=270]{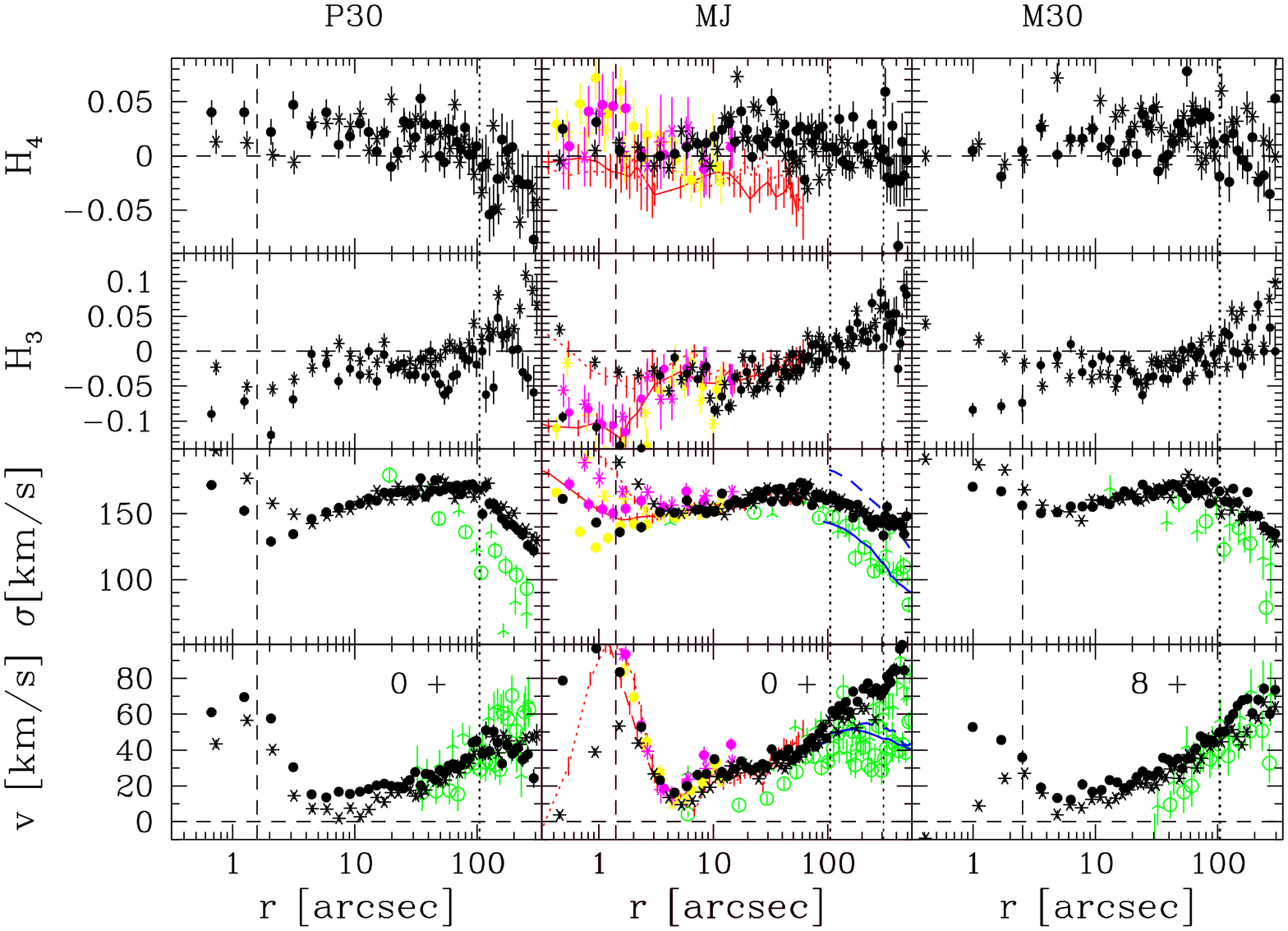}
\caption{The stellar kinematics along different position angles,
  folded with respect to the center (at velocity -333.2 km/s),
  antisymmetrically for recessional velocities and $H_3$ coefficients,
  symmetrically for velocity dispersions and $H_4$ coefficients.
  Filled circles refer to radii with increasing RA (East), stars to
  radii with decreasing RA (West). The number in the velocity panel
  indicates the correction in km/s to the central velocity (-333.2
  km/s) applied to achieve maximal symmetry. The plus or minus sign
  indicates whether the velocities of the sides with increasing RA
  (the filled dots) are receding or approaching. The vertical dotted
  lines mark the transition from the central slit data to those from
  the outer two. For the major axis, a further vertical dotted line
  marks the outer MJEE dataset. The short-dashed vertical lines mark
  the seeing values of the central slits (see Table \ref{tab_log}).
  The data of McElroy (\cite{McElroy83}, at PA=75, 45, 15, 165, 135
  and 105 respectively) are indicated by green open circles and
  triangles.  The data of van der Marel et al.  (\cite{vdMarel94}, MJ
  at PA=55, MN at PA=128) are shown by the red continuous lines. The
  data of Kormendy and Bender (\cite{Kormendy99}) are shown in yellow
  (from their Calcium Triplet spectra) and magenta (from their
  H$\beta$-Mg-Fe spectra). The blue dashed lines show Model K1 of
  Widrow et al.  (\cite{Widrow03}), the blue full lines their Model A
  (see discussion in Sect.  \ref{sec_dyn}).}
\label{fig_kin}
\end{figure*}

\addtocounter{figure}{-1}
\begin{figure*}%[t!]
\centering
\includegraphics[width=14cm,angle=270]{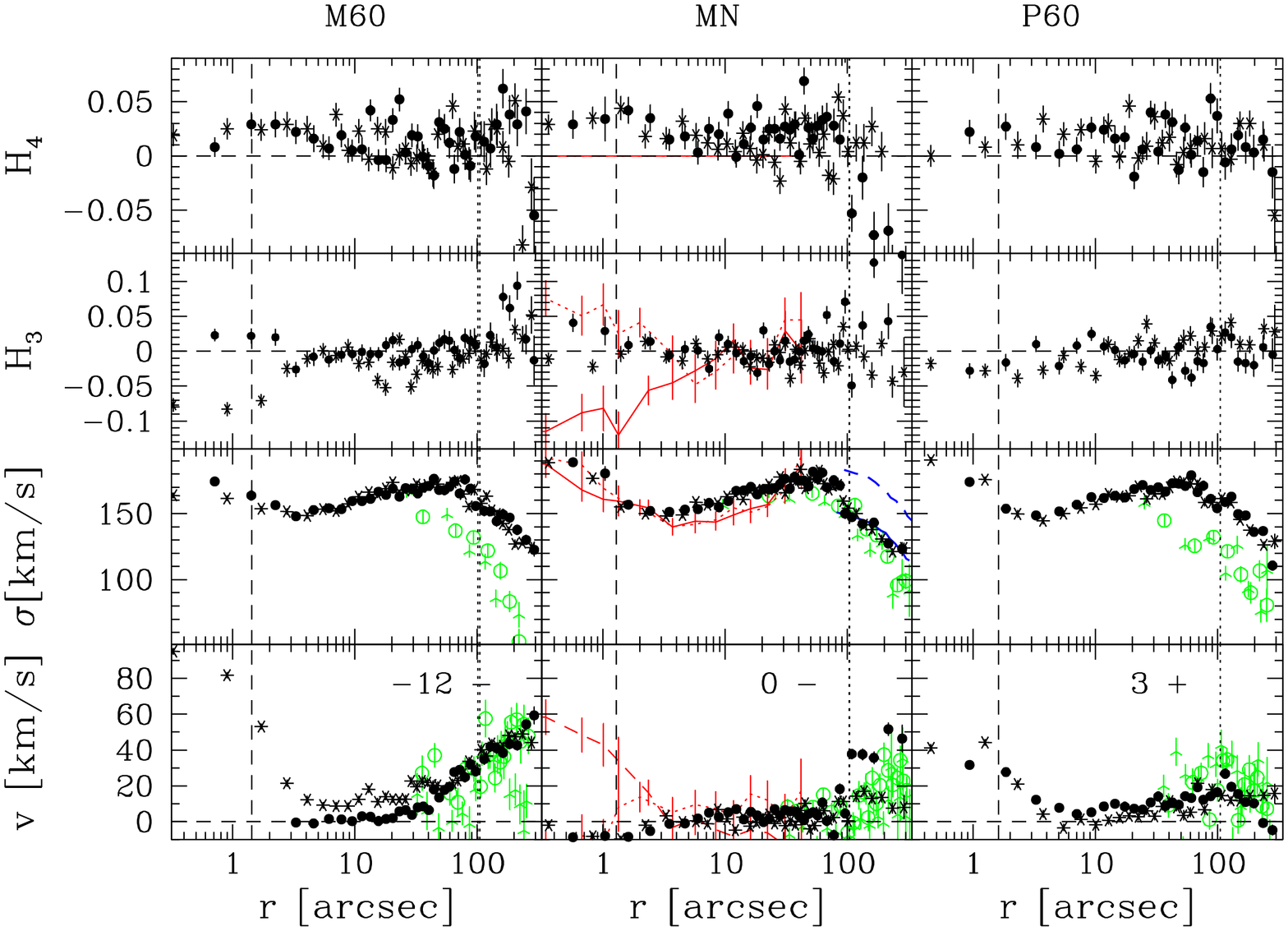}
\caption{Continued}
\end{figure*}

\begin{figure}[h!]
\centering
\includegraphics[width=7.2cm,angle=0]{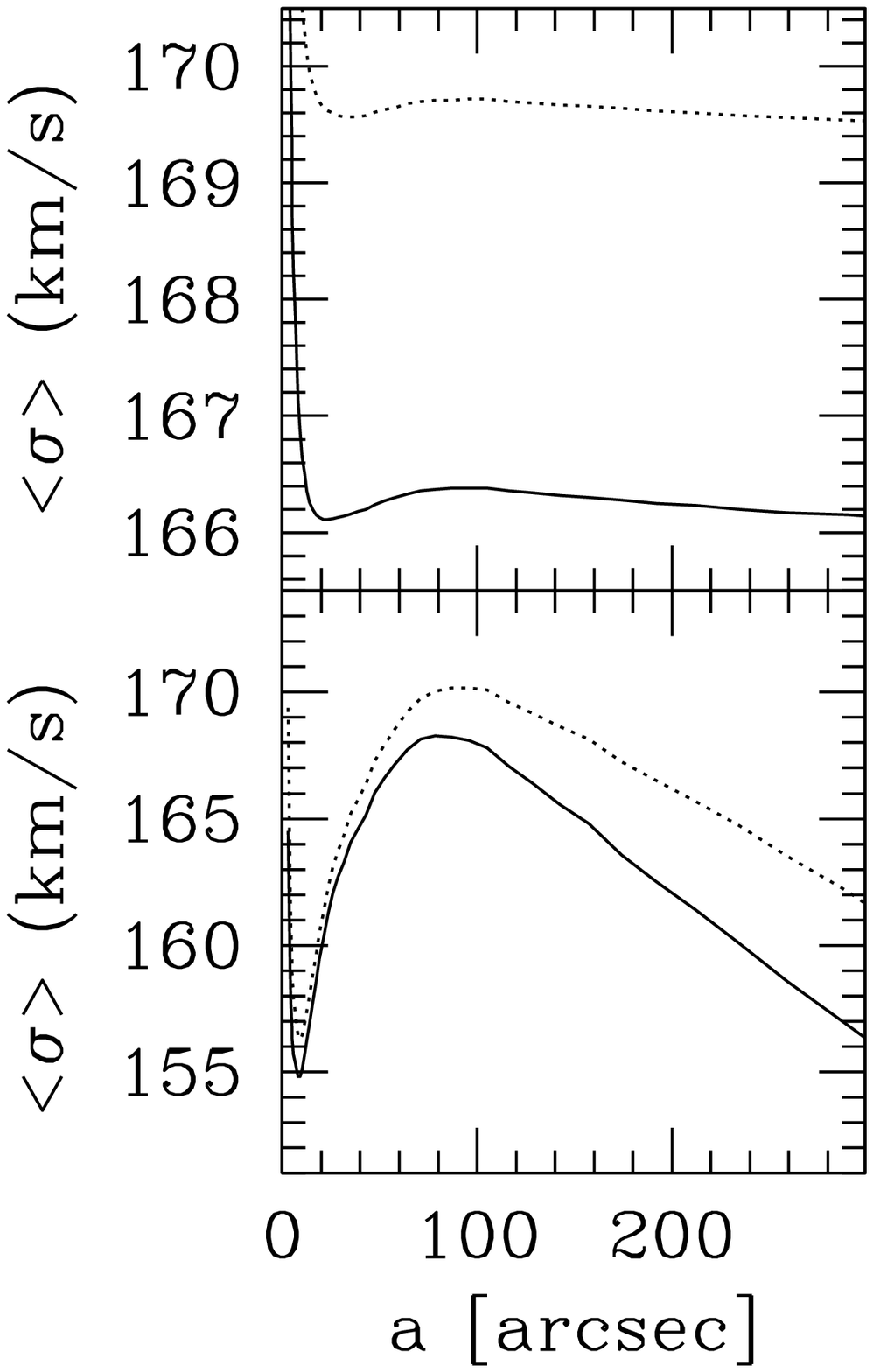}
\caption{Top: the full line shows the velocity dispersion resulting by
  averaging all measured $\sigma$ within the isophote with semi-major
  axis $a$ in arcsec, with weights equal to the surface luminosity at
  each point. The dotted line shows the same for the quantity
  $\sqrt{\sigma^2+V^2}$, where $V$ is the measured line of sight
  velocity. Bottom: the same computed by weighting according to the
  total isophotal light. }
\label{fig_meansigma}
\end{figure}

As a consequence, the mean velocity dispersion used to predict the
mass of the black hole at the center of M31 (160 km/s, G\"ultekin et
al. \cite{Gueltekin09}, corresponding to $M_{BH}=0.5\times
10^8M_\odot$, 3 times smaller than the measured value) has to be
revised upwards.  Fig.  \ref{fig_meansigma}, top, shows that the
(angle averaged) light-weighted velocity dispersion converges to 166
km/s and to almost 170 km/s if the rotational velocity is added in
quadrature. This results in a black hole mass only a factor 2 smaller
than observed.  The same figure, bottom, shows that lower values are
derived as a function of the distance from the center, if the
weighting is performed using the total isophotal light.

\begin{figure*}%[t!]
\centering
\includegraphics[width=15cm,angle=270]{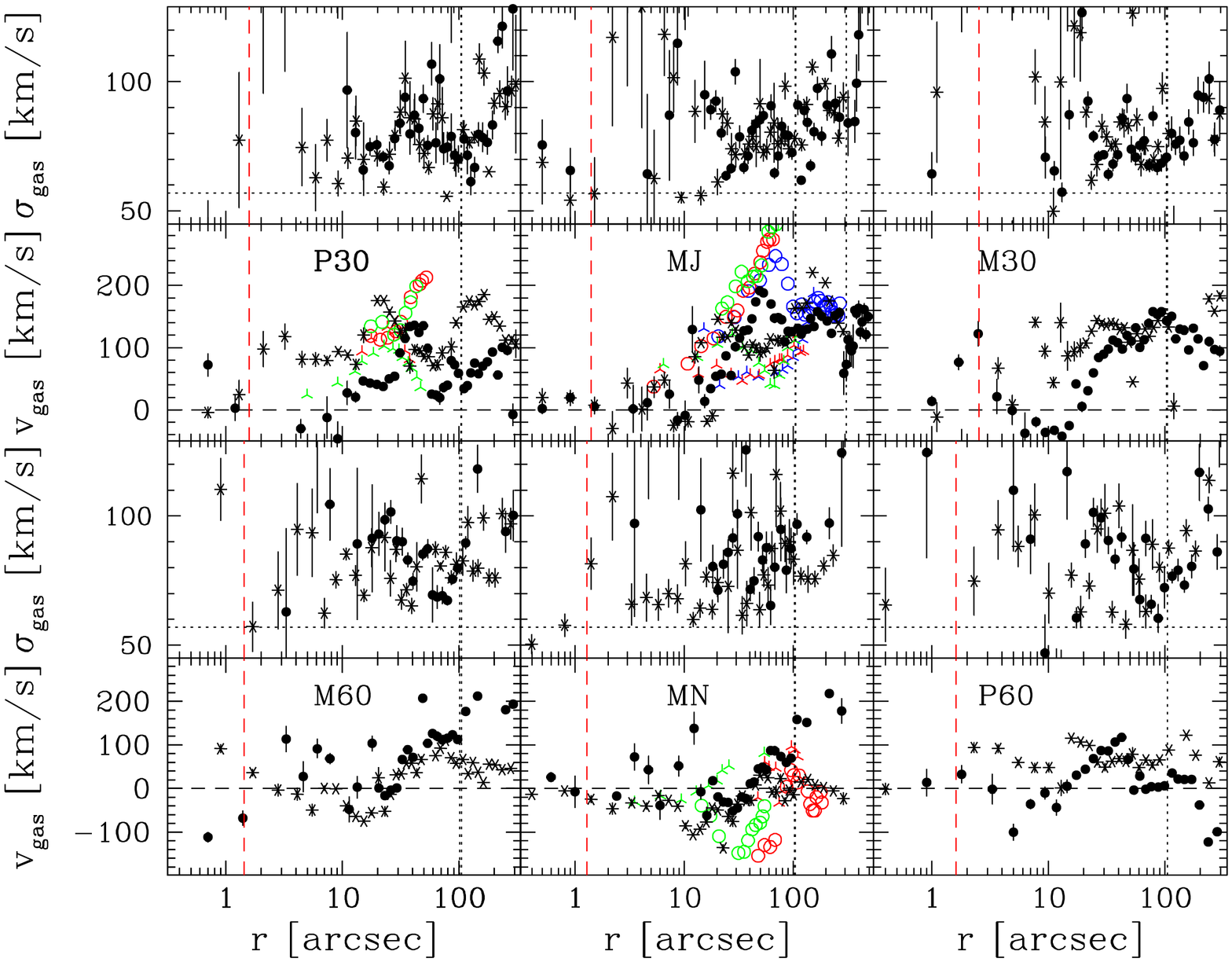}
\caption{The gas kinematics along different position angles, folded
  with respect to the center, antisymmetrically for recessional
  velocities, symmetrically for velocity dispersions. Only data with
  errors smaller than 50 km/s are shown. Filled circles refer to radii
  with increasing RA, stars to radii with decreasing RA, with the same
  convention and zero-point shifts on radial velocities of Fig.
  \ref{fig_kin}.  The vertical dotted lines mark the transition from
  the central slit data to those from the outer two.  For the major
  axis, a further vertical dotted line marks the outer MJEE dataset.
  The (red) short-dashed vertical lines mark the seeing values of the
  central slits (see Table \ref{tab_log}). The data of Pellet
  \cite{Pellet76}) are also shown for comparison as open circles (for
  increasing RA) and triangles (for decreasing RA).  His 28 \AA\ data
  are plotted green, the 135 \AA\ red.  We show his PA=45 data as MJ,
  his PA=128 as MN, his PA=68 as P30.  Moreover, the data at PA=38
  (his major axis) are shown blue also as MJ.}
\label{fig_kingas}
\end{figure*}

The gas kinematics ($V_{gas}$ and $\sigma_{gas}$) are shown in Fig.
\ref{fig_kingas}. Gas velocities are almost a factor two higher than
stellar velocities, but less axisymmetric and regular. Gas velocity
dispersions are slightly larger than the instrumental resolution,
indicating intrinsic velocity dispersions less than 80 km/s. In the
inner 10-20 arcsec near the minor axis gas counterrotation (i.e., the
orientation of the gas rotation changes) is observed. The gas
velocities compare reasonably well with the H$\alpha$ measurements of
Pellet (\cite{Pellet76}) or Boulesteix et al.  (\cite{Boulesteix87}).

We measured Lick line strength index profiles on the M31
emission-corrected spectra as in Mehlert et al. (\cite{Mehlert00}),
from H$\beta$ to Fe5789. 

\begin{figure}%[h!]
\centering
\includegraphics[width=8cm]{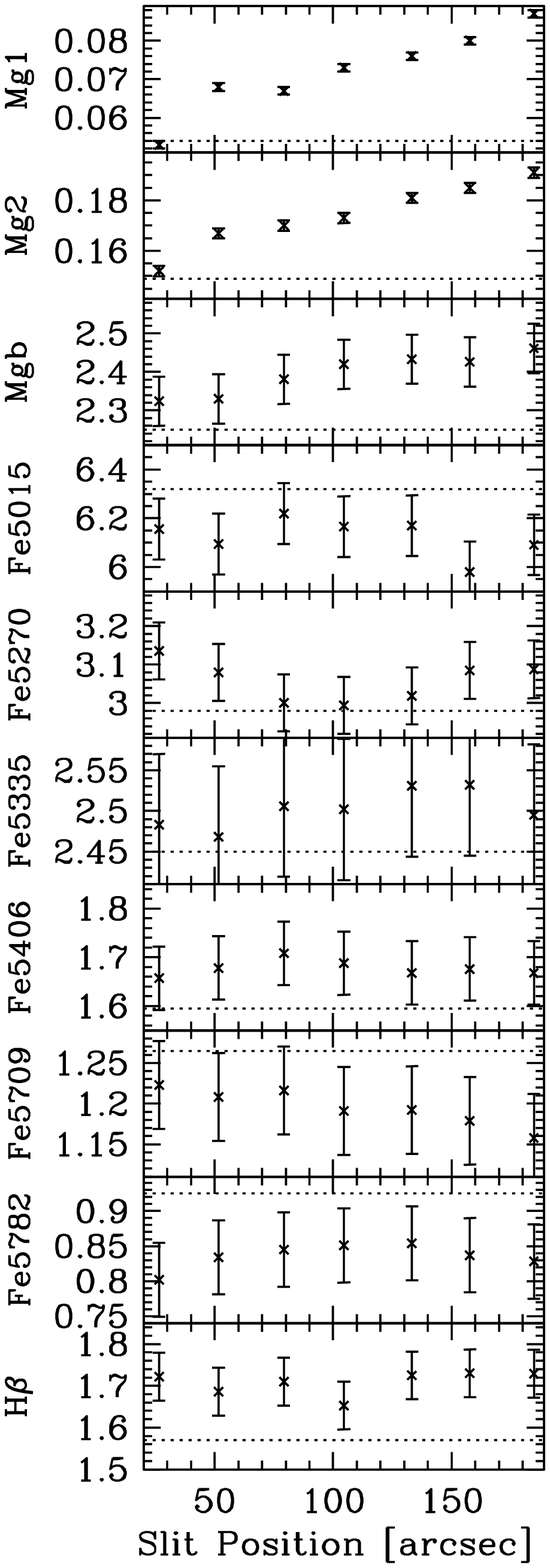}
\caption{The variation of the molecular Mg1 and Mg2 (in mag) and atomic 
(in \AA , all the
  rest) line strengths as a function of position on the slit measured
  for the star HD72324 (= HR 3369). The dotted line shows the values of Worthey
  et al. (\cite{Worthey94}).}
\label{fig_lickstarprofile}
\end{figure}

As a first check, we verify that we measure line indices uniformly
well across the slit. For this purpose, we analyse the spectra of the
Lick standard star HD 72324 (= HR 3369) that was observed trailed along the
slit.  Fig. \ref{fig_lickstarprofile} shows the variation of its line
indices as a function of the position on the slit.  The measured
molecular indices Mg1 and Mg2 with their widely spread continuum
windows are increasingly biased towards the ends of the slit, probably
because of inaccurate spectral flat fielding where vignetting becomes
important.  In contrast, all the other indices are well determined and
do not vary much . In the following we do not consider Mg1 and Mg2
anymore.  As a consequence, all (atomic) indices presented here below
are measured in Angstrom.

%\clearpage

\begin{figure*}%[h!]
\centering
\includegraphics[width=15.cm]{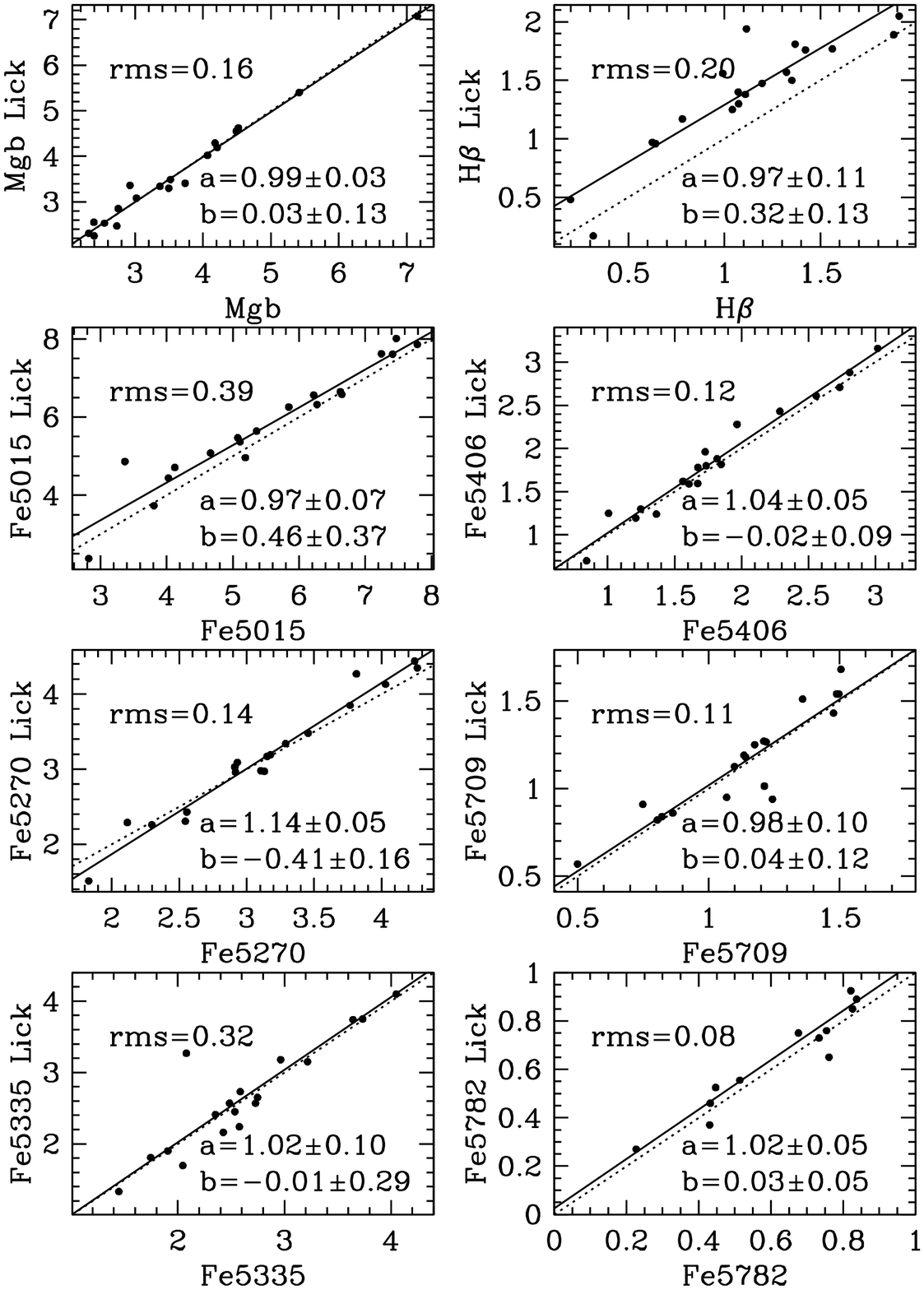}
\caption{The comparison between the line strengths measured here and
  the Lick values for 20 stars. The dotted line shows the one-to-one
  correlation, the full line the best fitting relation
  $I(Lick)=a\times Lick(Us)+b$.}
\label{fig_lick}
\end{figure*}

As a second step, we assess how well we are on the Lick system,
since our spectra are not flux calibrated. Our indices are measured
following the bands definitions of Worthey et al. (\cite{Worthey94})
with the corrections of Trager et al. (\cite{Trager98}).  Fig.
\ref{fig_lick} demonstrates that no tilt nor shift are needed within
the errors, with a typical rms of 0.2 \AA .  The largest shift, 0.32
\AA , significant at the 2.5 sigma level, is observed for the H$\beta$
line, where, as discussed above, we have interpolated two bad columns.
We have added in quadrature the rms of each relation to the errors of
the galaxy indices. 

\begin{figure}%[h!]
\centering
\includegraphics[width=8cm]{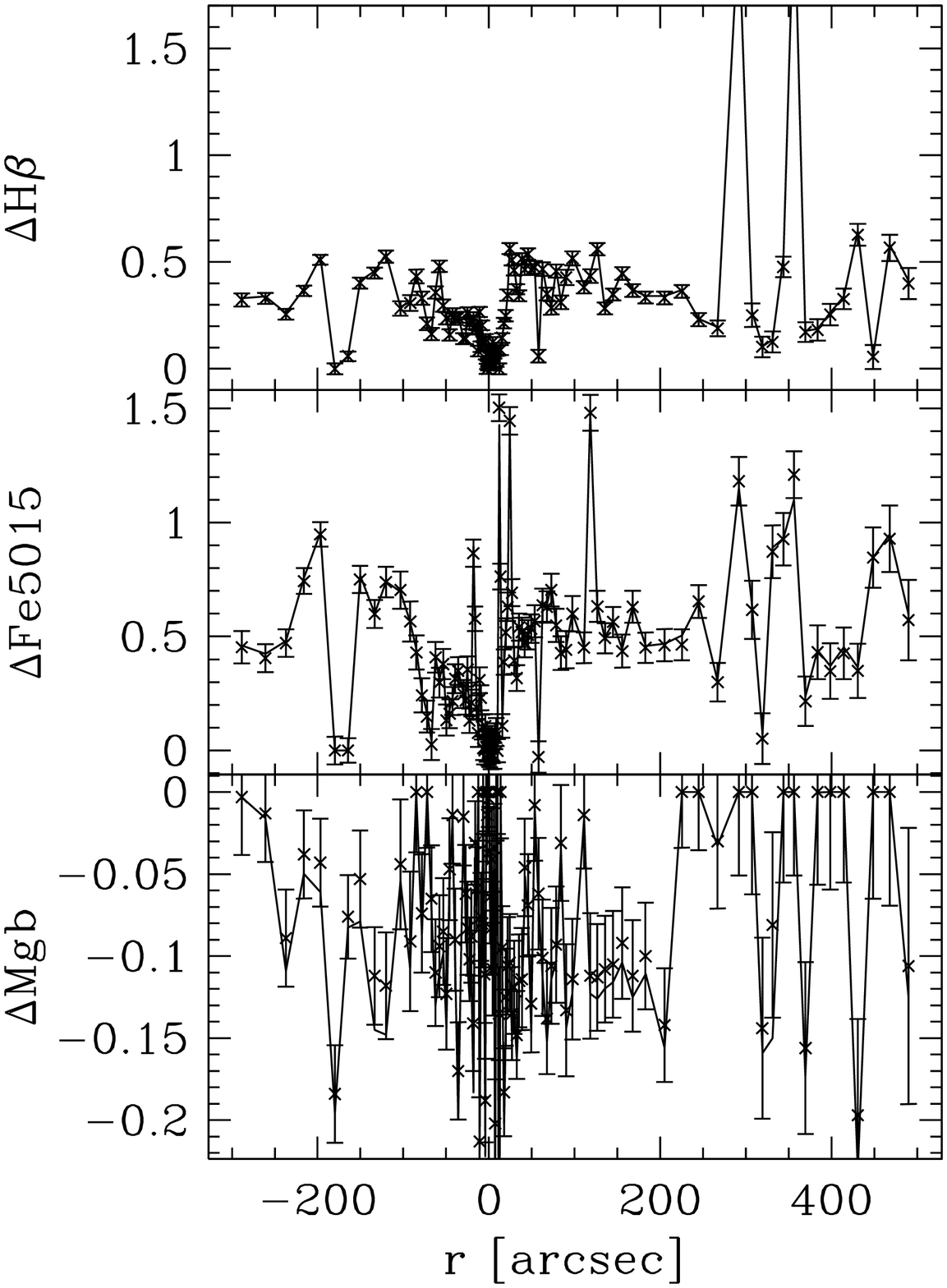}
\caption{The effect of the gas emission correction on the H$\beta$,
  Fe5015 and Mgb indices along the major axis of M31. The plot shows
  the differences $\Delta I=I_{cor}-I_{raw}$ between the
  emission corrected index $I_{cor}$ and the values $I_{raw}$ measured
  before the correction. The full lines show the EW of the measured
  emissions in H$\beta$ and [NI] and the difference between the EWs of 
  the redder and bluer emission lines of the [OIII] doublet (see text).}
\label{fig_indexgas}
\end{figure}

Finally, Fig.  \ref{fig_indexgas} shows the typical effect of the
correction for gas emission applied to the spectra. The plot shows the
differences $\Delta I=I_{cor}-I_{raw}$ between the
emission corrected index $I_{cor}$ and the values $I_{raw}$ measured
before the correction. Before the correction, the H$\beta$ and Fe5015
indices are smaller (since the emission is present in the index
window) by typically 0.29 \AA\ and 0.41 \AA\ respectively, and Mgb is
larger (since the emission is present in the redder continuum window)
by 0.07 \AA. While the $\Delta$H$\beta$ and $\Delta$[NI] match fully
the EW of the emission, $\Delta$[OIII] amounts to the difference
between the EWs of the [OIII] doublet redder line (that falls in the
band window of the Fe5015 index) and the bluer line (that falls in the
bluer continuum window of the index). The other indices are not
affected. Again, we add in quadrature the errors on the emission EWs
of H$\beta$, [OIII] and [NI] to the ones of the absorption line
strengths H$\beta$, Fe5015 and Mgb, respectively.

Fig. \ref{fig_lickprofiles} shows the measured Lick indices profiles
for M31.  Apart from the central regions, where rapid variations are
measured, the indices show very mild gradients, indicative of
homogeneous stellar populations (see discussion in Sect.
\ref{sec_stelpop}). Along the major axis and at distances larger than
$\approx 5$ arcmin, where the disk light starts to dominate, making
55\% of the total at 500 arcsec from the center, the strenght of the
$H\beta$ line increases, indicative of a younger stellar population
(see discussion in Sect.  \ref{sec_stelpop}).  Overall, the indices
appear not only symmetric with respect to the center, but also
approximately symmetric on {\it circles}. As an example, if we compute
the rms of the differences of the Mgb index measured at the same
distance on the major axis and on one of the other slit positions, we
get values between 0.15 and 0.18 \AA . If we now repeat the procedure
interpolating at the same {\it isophotal} distance (using the
ellipticity profile derived in Sect.  \ref{sec_dyn}, see Fig.
\ref{fig_triaxial}), the rms increases to 0.20-0.25 \AA .  The same
applies to the H$\beta$ or the Iron indices.

Davidge (\cite{Davidge97}) measured Lick indices (not corrected for
gas emission) in a 22$\times$60 arcsec area centered on M31. His
values along the centered East-West slit are shown in
Fig. \ref{fig_indexgas}, on the plots for the P30 slit 
(nearest to the EW position). The agreement is good.

\begin{figure*}%[t!]
\centering
\includegraphics[width=15cm]{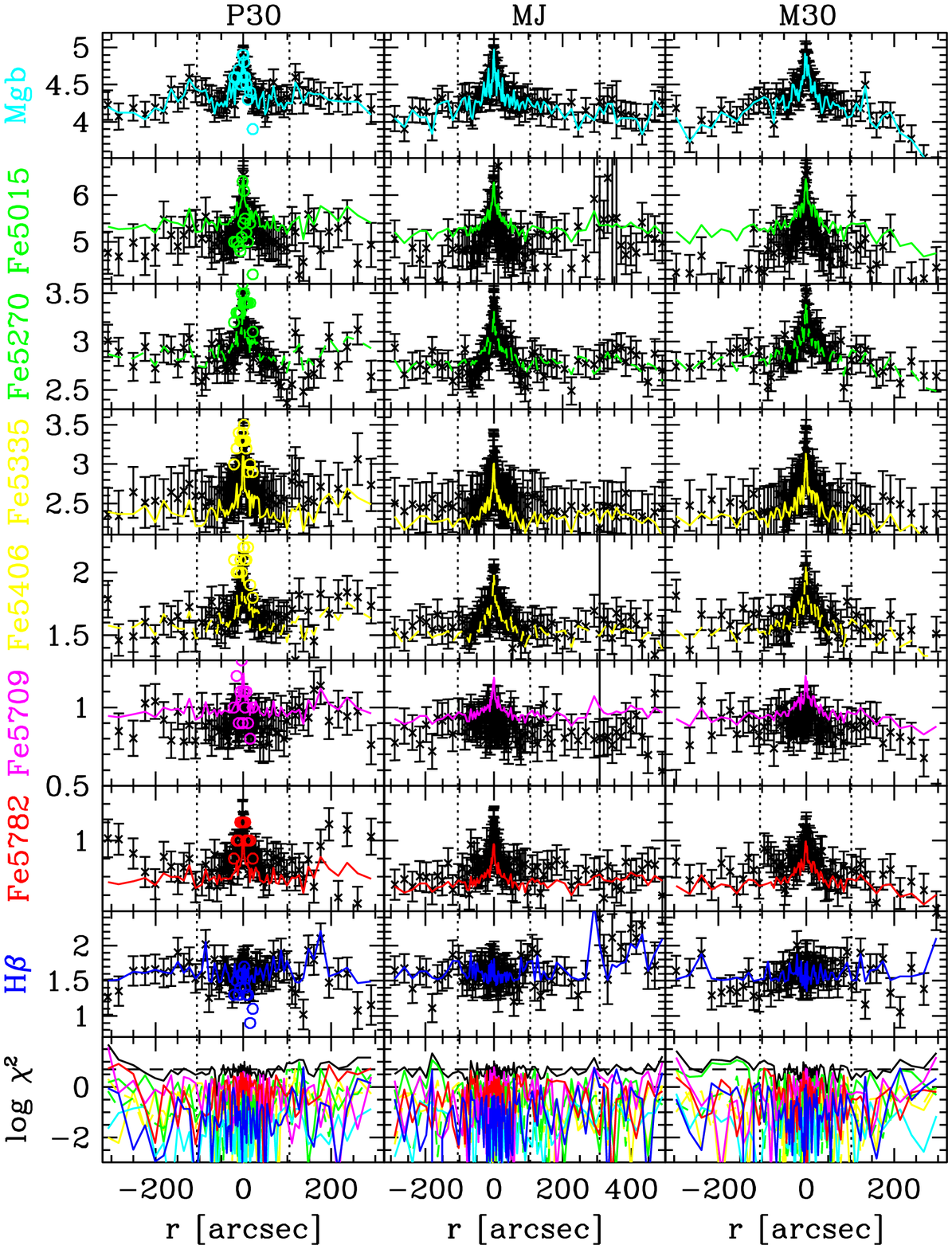}
\caption{The measured Lick indices profiles along the 6 slit positions
  of M31.  The vertical dotted lines mark the ends of the central
  slits.  For the major axis, a further vertical dotted line marks the
  outer MJEE dataset. The lines show the best-fit LSSP models (see
  discussion in Sect.  \ref{sec_stelpop}). The bottom plots show the
  respective $\log \chi^2$, coded by color, and the total $\log
  \chi^2$ in black. The long-dashed line shows the expected value for
  the given degrees of freedom (5). The open symbols in P30 show the
  data of Davidge (\cite{Davidge97}). \label{fig_lickprofiles}}
\end{figure*}

\addtocounter{figure}{-1}
\begin{figure*}%[t!]
\centering
\includegraphics[width=15cm]{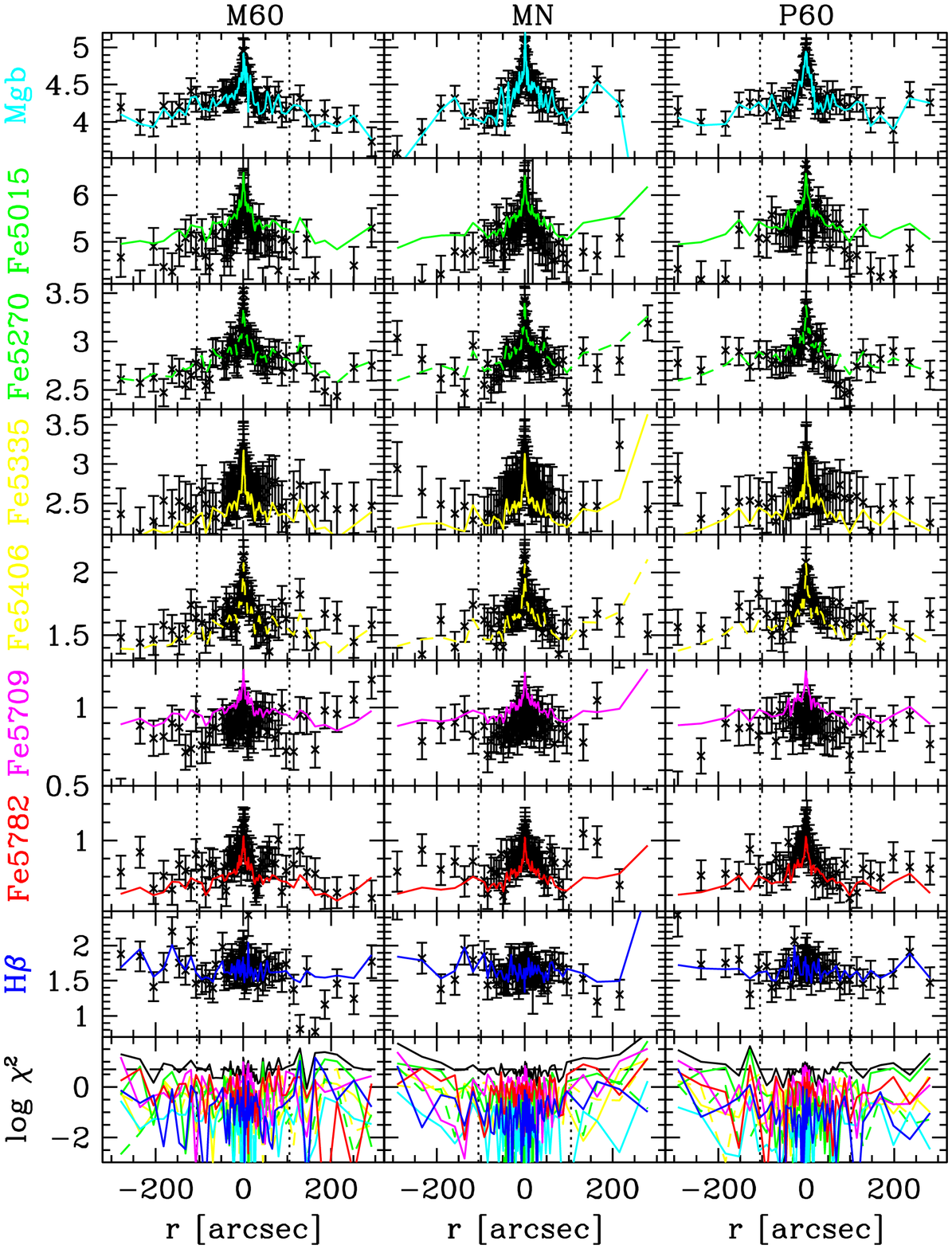}
\caption{Continued.}
\end{figure*}

\begin{figure*}%[h!]
\centering
\includegraphics[width=15cm,angle=270]{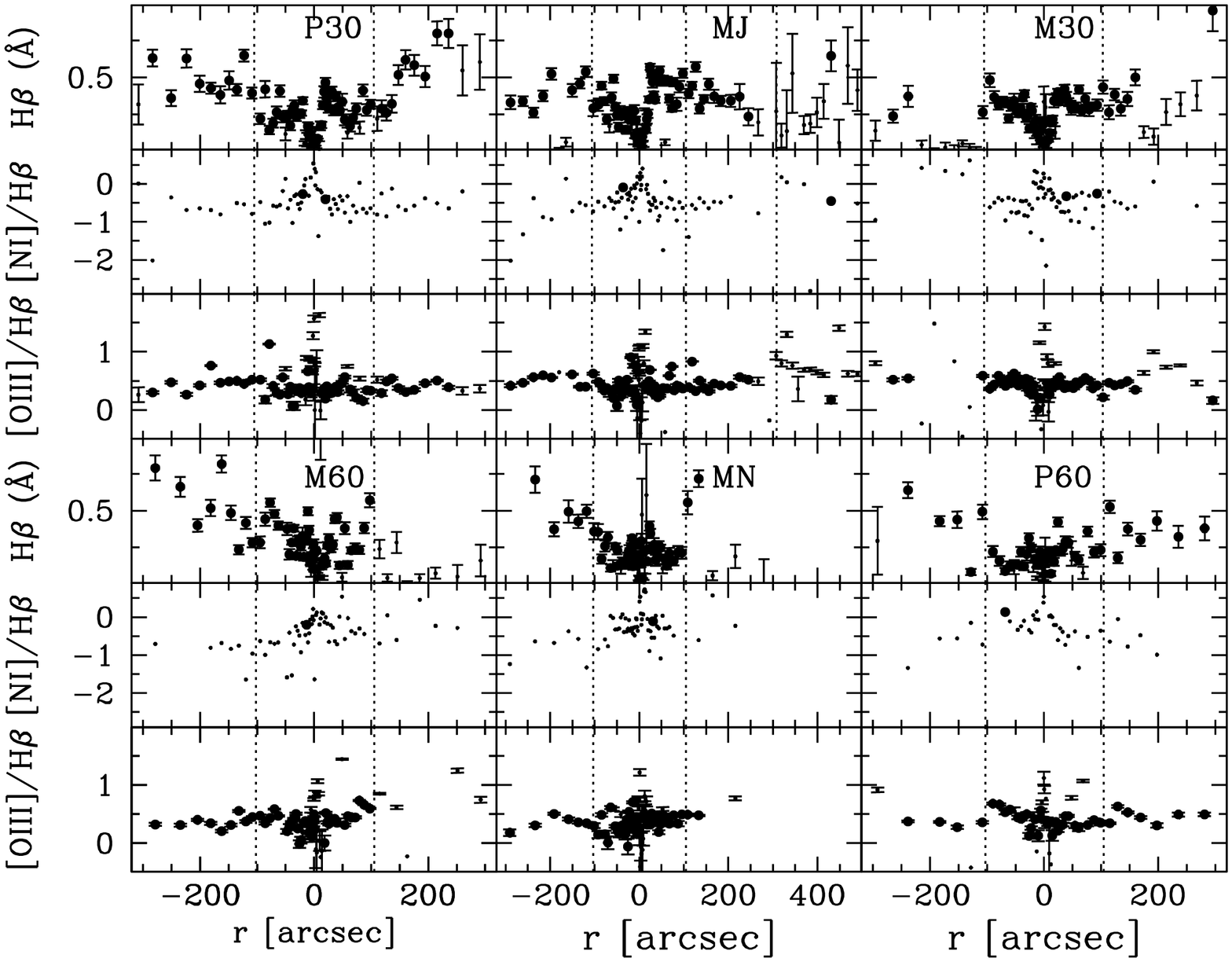}
\caption{The EW of the H$\beta$ emission line in \AA\ and the ratios
  [OIII]/H$\beta$ and [NI]/H$\beta$ of the emission EWs in logarithmic
  units as a function of radius along different position angles. The
  large filled circles show the points with signal to noise ratio
  larger than 4. The vertical dotted lines mark the ends of the
  central slits. For the major axis, a further vertical dotted line marks the
  outer MJEE dataset.\label{fig_gasratios} }
\end{figure*}

In Fig. \ref{fig_gasratios} we plot the equivalent widths (EWs) of the
H$\beta$ emission line in \AA , together with the EW ratios [OIII]/H$\beta$
and [NI]/H$\beta$ of the emission lines considered above in
logarithmic units. Note that we expect the EW ratios to vary by no
more than 10\% with respect to the ratios of the emission line fluxes,
due to the restricted wavelength range spanned by the three lines.
Given the logarithmic units, this does not affect the discussion given
in Sect. \ref{sec_ionize}. While the
H$\beta$ and [OIII] lines are almost always well detected, with a
signal to noise ratio larger than 4, the [NI] doublets are often
weak, if not undetected. The H$\beta$ EW is approximately constant
with radius.  At radii larger than $\approx 10$ arcsec the same is
true for the [OIII]/H$\beta$ ratio (at the value of 3) and the
[NI]/H$\beta$ ratio (at low values smaller than $\approx 0.1$).  Near
the center [OIII]/H$\beta$ increases to $\approx 10$, combined with an
increased [NI]/H$\beta$ ratio ($\approx 1$).

Tables \ref{tab_stellardata}, \ref{tab_gasdata}, and \ref{tab_Lickdata} 
give format examples of the measured stellar and gaseous kinematics,
and Lick indices as a function of distance and position angle, respectively.
The full listing is available electronically.

\begin{table*}
\caption{Format example of the measured stellar kinematics as a function of distance from the center (positive: east, negative: 
west) for the different position 
angles. The full table is available electronically.} 
\label{tab_stellardata}
\begin{tabular}{cccccc}
\hline
R & PA & $V_{stars}$ & $\sigma_{stars}$ & $H_3$ & $H_4$ \\ 
(``) & & (km/s) & (km/s) & & \\
\hline
-289 &  MJ & $-404.14\pm  2.21$ &  $143.23\pm 2.60$ &  $-0.061\pm 0.014$ & $-0.002\pm 0.014$   \\
\hline
\end{tabular}
\end{table*}

\begin{table*}
\caption{Format example of the measured gas kinematics as a function of distance from the center (positive: east, negative: 
west) for the different position 
angles. The full table is available electronically.} 
\label{tab_gasdata}
\begin{tabular}{ccccccccc}
\hline
R & PA & $V_{gas}$ & $\sigma_{gas}$ & H$\beta$ & [OIII]/H$\beta$ & [NI]/H$\beta$ \\ 
(``) & & (km/s) & (km/s) & \AA & & \\
\hline
-289 &  MJ & $-458.7\pm 4.9$ &  $94\pm5.6$ & $0.33\pm0.046$ & $2.61\pm0.15$ & $0.01\pm 15.7$\\
\hline
\end{tabular}
\end{table*}

\begin{table*}
\caption{Format example of the measured Lick indices as a function of distance from the center (positive: east, negative: 
west) for the different position 
angles. The full table is available electronically.} 
\label{tab_Lickdata}
\begin{tabular}{llllllllll}
\hline
R & PA & H$\beta$ & Mgb & Fe5015 & Fe5270 & Fe5335 & Fe5406 & Fe5709 & Fe5782 \\
(``) & & \AA & \AA & \AA &\AA &\AA &\AA &\AA &\AA  \\
\hline
-289 & MJ & $1.65\pm 0.21$ & $4.04\pm 0.17$ & $5.07\pm 0.40$ & $2.94\pm 0.14$ & $2.50\pm 0.32$ & $1.53\pm 0.12$ & $0.78\pm 0.11$ & $0.78\pm 0.08$\\
\hline
\end{tabular}
\end{table*}

\section{An old and massive bulge?}
\label{sec_mod}

Before discussing the stellar populations (Sect. \ref{sec_stelpop})
and dynamics (Sect. \ref{sec_dyn}) of the central 300 arcsec of M31,
it is important to assess what is the light contribution of the bulge
and the disk of M31 as a function of distance and position angle.
According to the decomposition of Kormendy and Bender
(\cite{Kormendy99}), based on a Sersic plus exponential law fit to the
MJ photometry and very similar to the one of Kent (\cite{Kent89b}),
the disk makes 1\% of the total light at the center, 10\% at 100
arcsec and 31\% at a distance of 300 arcsec along the major axis, and
9\% and 25\% along the minor axis respectively. The disk light amounts
to < 55\% of the total at the end of the MJEE slit, 500 arcsec from
the center.  The results discussed in Sect.  \ref{sec_stelpop} show
that generally old stellar populations are found out to 300 arcsec
from the center. In agreement with Stephens et al.
(\cite{Stephens03}), who start detecting (younger) disk stars only at
distances larger than 7 arcmin from the center, we see the effect of
the increasing importance of the younger disk component only at
distances of 400-500 arcsec from the center. In contrast, the
kinematic imprint of the disk stars is probably visible already at
distances $\ge 100$ arcsec, especially on the major axis (see Fig.
\ref{fig_kin} and Sect.  \ref{sec_kinematics}). We discuss it in Sect.
\ref{sec_dyn}.

\subsection{Stellar populations}
\label{sec_stelpop}

We study the stellar populations of the inner 5 arcmin of M31 
(and out to 8 arcmin along the major axis) using
the simple stellar population (SSP) models of Maraston
(\cite{Maraston98}, \cite{Maraston05}) with a Kroupa (\cite{Kroupa01})
IMF, and the modeling of the Lick indices (LSSP) with
$\alpha-$elements overabundance of Thomas, Maraston and Bender
(\cite{TMB03}). The models cover ages $t$ up 15 Gyr, metallicities
[Z/H] from -2.25 dex to +0.67, and overabundances [$\alpha$/Fe] from
-0.3 to +0.5. We spline-interpolate the LSSP models on a finer grid in
$t$, [Z/H] and [$\alpha$/Fe] and
determine values of $t$, [Z/H] and [$\alpha$/Fe] that minimize the
$\chi (r)^2$:
\begin{equation}
\label{eq_chi}
\begin{array}{ccl}
\chi (r)^2& = &\Delta H\beta(r)^2+\Delta Mgb(r)^2+\Delta Fe5015(r)^2+\\
         &   &\Delta Fe5270(r)^2+\Delta Fe5335(r)^2+\Delta Fe5406(r)^2+\\
         &   & \Delta Fe5709(r)^2+\Delta Fe5782(r)^2\\
\end{array}
\end{equation} 
at each radius $r$, where $\Delta Index(r)^2$ is:
\begin{equation}
\label{eq_delta2}
\Delta Index (r)^2=\left(\frac{Index(r)-mIndex(t,[Z/H],[\alpha/Fe])}
{dIndex(r)}\right)^2,
\end{equation}
where $dIndex(r)$ is the error on the index $Index(r)$ measured at
distance $r$ and $mIndex(t,[Z/H],[\alpha/Fe])$ is the value predicted
by the LSSP models for the given set of age, metallicity and
overabundance. Note that we do not extrapolate to values outside the
model grid. Therefore the minimum $\chi^2$ solutions are sometimes
found at the edges of the parameter space. Once the set
$(t_{min}(r),[Z/H]_{min}(r),[\alpha/Fe]_{min}(r))$ that minimizes
$\chi(r)^2$ has been determined, we compute the values of the colors
and mass-to-light (M/L) ratios of the corresponding SSP models. Errors
on all quantities are derived by considering the minimal and maximal
parameter variations compatible with $\Delta \chi ^2 = \chi ^2-\chi
^2_{min}=1$.

Fig. \ref{fig_SPPmodels} shows the resulting age, metallicity,
overabundance, color and M/L profiles along the different slit
positions. The corresponding best-fit lines to the measured indices
profiles are shown in Fig. \ref{fig_lickprofiles}. We fit 8 Lick
indices to derive three parameters, so the expected $\chi^2$ should be
around 5. This matches rather well the derived values (see bottom plot
of Fig. \ref{fig_lickprofiles}), indicating that the SSP models are a
good representation of the data within their sizeable errors.

The six age, metallicity and overabundance slit
profiles are very similar when overplotted as a function of radius.
This is a consequence of the approximately constancy of the indices
along circles discussed above and indicates that the region with
homogeneous stellar populations is not as flattened as a disk
component would be, in agreement with what found by Davidge
(\cite{Davidge97}) for the inner 30 arcsec.
Fig.  \ref{fig_aveSSP} shows the age, metallicity,
overabundance, and $M/L_R$ profiles averaged over position angles and
binned logarithmically in radius (for the MN profile, only the central slit
data are included in the average).

\begin{figure*}%[t!]
\centering
\includegraphics[width=15cm]{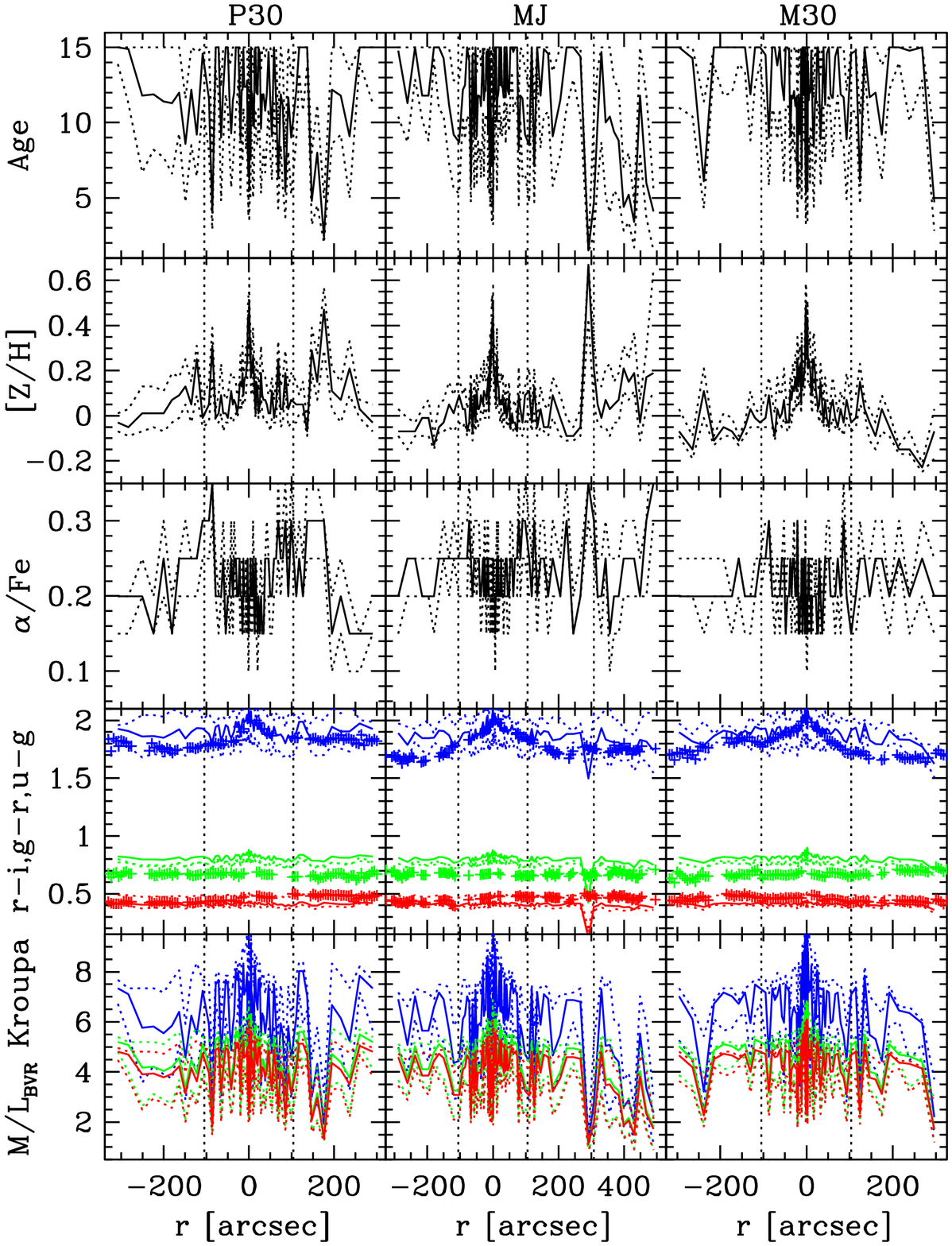}
\caption{The age, metallicity, overabundance, Sloan colors u-g (blue),
  g-r (green), r-i (red) and Johnson M/L profiles in the B (blue), V
  (green) and R (red) bands along the different slit positions. The
  dotted lines show the 1-sigma errors. The vertical dotted lines mark
  the ends of the central slits.  For the major axis, a further
  vertical dotted line marks the outer MJEE dataset. The measured
  color profiles, corrected by extinction (Montalto et al.
  \cite{Montalto09}) are shown as crosses.}
\label{fig_SPPmodels}
\end{figure*}

\addtocounter{figure}{-1}
\begin{figure*}%[t!]
\centering
\includegraphics[width=15cm]{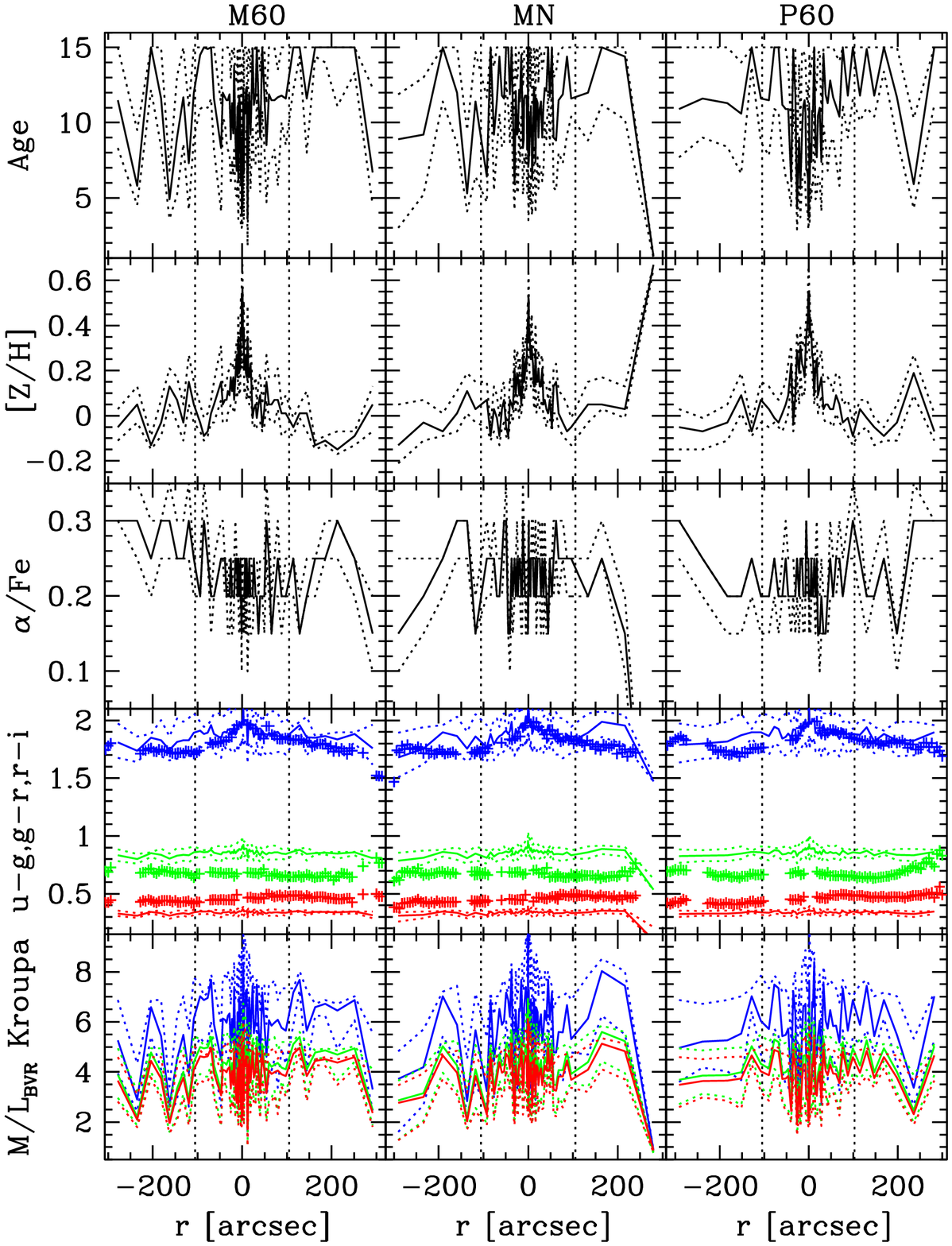}
\caption{Continued.}
\end{figure*}

\begin{figure}%[t!]
\centering
\includegraphics[width=8cm]{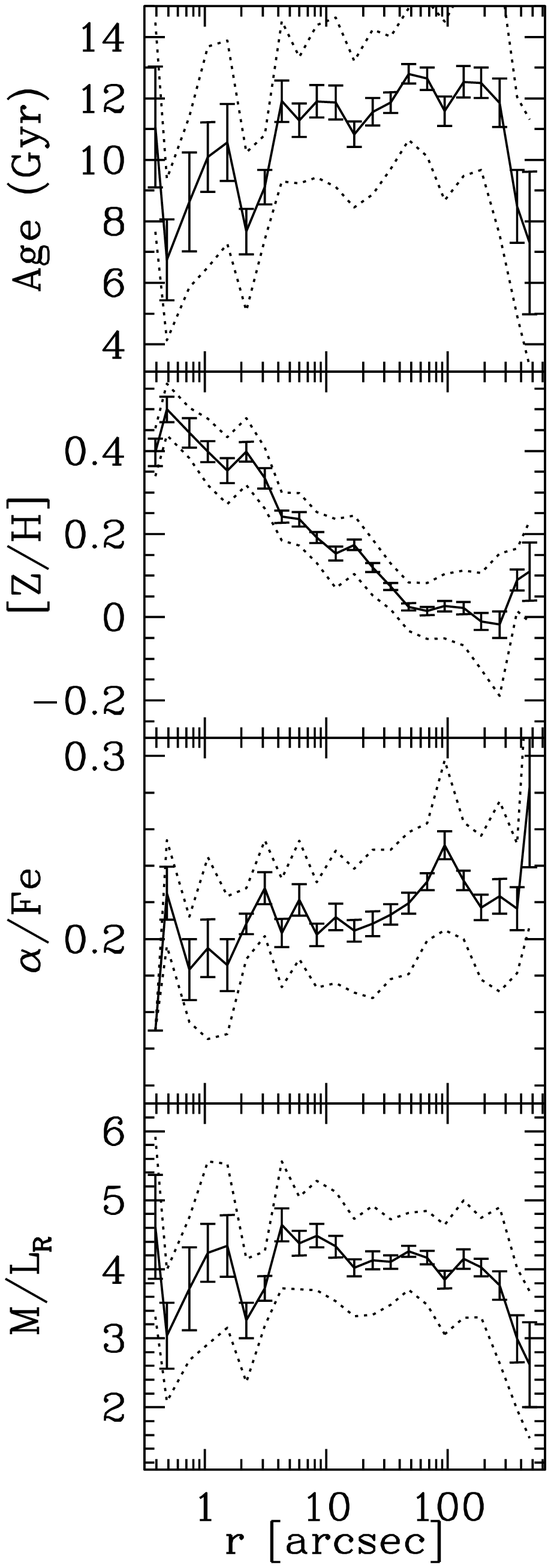}
\caption{The age, metallicity, overabundance, and $M/L_R$ profiles
  (full lines) averaged over position angles and binned
  logarithmically in radius. The dotted lines and the bars show the
  corresponding rms and errors on the mean, respectively.
  \label{fig_aveSSP}}
\end{figure}

The picture emerging from Fig. \ref{fig_SPPmodels} and
\ref{fig_aveSSP} is the following.  At radii larger than a few arcsec
and out to $\approx 5$ arcmin 
from the center, the stars of the bulge of M31 are on average almost
as old as the universe, tend to solar metallicity and are $\alpha$
overabundant by 0.2-0.25 dex.  This is in agreement with Stephens et
al. (\cite{Stephens03}), who find that the near infrared luminosity
function of the red giant stars of the bulge of M31 is
indistinguishable from the one of the Milky Way bulge. Of course, the
point to point scatter in age in the single slit profiles is large and
formally the 1-sigma lines would allow sometimes ages as small as 5
Gyr. This is driven by the scatter in the H$\beta$, that ultimately
correlates with the emission correction applied.  There is a
metallicity gradient of 0.2 dex per radial decade out to 40 arcsec,
and essentially no gradient at larger distances, where solar
metallicity is reached. There is no obvious radial variation in age or
overabundance. This translates in moderate u-g and M/L$_B$ gradients,
and almost no detectable gradients in the g-r, r-i, M/L$_V$ and
M/L$_R$. The exact values of the stellar mass-to-light ratios depend
on the assumed IMF (Kroupa \cite{Kroupa01}), that has a slope
$\alpha=1.3$ for the mass range $0.1<M<0.5 M_\odot$ and $\alpha=2.3$
at larger masses. Changing the break mass from $0.5
M_\odot$ to $1M_\odot$ lowers the $M/L_R$ from 4 to 3.

Along the major axis and at distances larger than $\approx 6$~arcmin,
the mean age of the stellar population drops to $\le 8$ Gyr, again in
agreement with Stephens et al. (\cite{Stephens03}), who find disk
stars only at distances larger than 7 arcmin from the center. The
metallicity is slightly supersolar (by $\approx +0.1$ dex). The
$\left[\alpha/Fe\right]$ overabunce remains high. As a consequence,
the derived mass-to-light ratio drops to $M/L_R\le 3M_\odot/L_\odot$.

The situation is different in the inner arcsecs, where
the stellar population age drops down to 4-8 Gyr, the metallicity
increases to above 3 times solar, but the overabundance remains at
$\approx 0.2$ dex. This is in broad agreement with the conclusions of
Davidge (\cite{Davidge97}), who however hints to a possible further
increase of the overabundance towards the center, not seen here.

We compare the predicted Sloan color profiles to the ones derived from
the Sloan survey, that have been dereddened using the extinction map
discussed in Montalto et al. (\cite{Montalto09}). Following Maraston
(private communication) we apply a correction of -0.05 mag to the
model $g-r$ and of +0.07 mag to the model r-i colors. The comparison
is surprisingly good in u-g and good in r-i, with deviations
$\le0.1$ mag. Some residual systematic deviations are still present in
g-r (where the models are too red) and r-i (where the models are slightly too
blue), in qualitative agreement with the discussion of Maraston et al.
(\cite{Maraston09}). This overall match provides an independent
assessment of the line index analysis.

\subsection{Axisymmetric dynamical modeling}
\label{sec_dyn}

The kinematics shown in Fig. \ref{fig_kin} describe the light averaged
stellar motions of the bulge and disk components. In the inner arcmin
this follows closely the bulge kinematics, but at larger distances the
contamination by the disk becomes more and more important. Therefore,
a dynamical model for the sum of the two components should be compared
to Fig. \ref{fig_kin}. Qualitatively, we can expect the stellar disk
to rotate much faster, possibly up to the 150 km/s observed in the gas
(see Fig. \ref{fig_kingas}), and to be colder (with velocity
dispersions of the order of 60-70 km/s) than the bulge. As a
consequence, the true rotational velocity of the bulge could be smaller
than, and the bulge velocity dispersion could be larger than shown in Fig.
\ref{fig_kin}. Moreover, the line of sight velocity distribution in
the outer regions will be dominated in the high velocity tail by disk
stars. This can balance the intrinsic asymmetry of the bulge (that
anticorrelates with the mean rotational velocity), possibly explaining the
change in sign of the $H_3$ coefficient observed near the major axis
at distances larger than 100 arcsec. We have tested that this scheme
could work with a simple bulge plus disk kinematical model that follows the
light decomposition of Kent (\cite{Kent89b}), but we postpone a
quantitative analysis to Morganti et al. (\cite{Morganti09}).

The blue lines in Fig. \ref{fig_kin} show the fits to the data of
McElroy (\cite{McElroy83}) by Widrow et al. (\cite{Widrow03}), model A
(full) and model K1 (dashed). From the discussion given above, 
it is clear that it is dangerous to assume that McElroy's data 
describe the kinematics of the bulge component alone, as done  
by Widrow et al. (\cite{Widrow03}). Indeed, McElroy (\cite{McElroy83})
recognises that he might be underestimating the true stellar velocity 
dispersion due to how his analysis method reacts to disk light contamination 
(see his Table 3), but concludes that the effect is not large 
on the basis of the disk to bulge decomposition available at the time, that 
severely downplayed the role of the disk in the inner 5 arcmin. 

Letting aside this intrinsic problem, the best-fit model A of Widrow
et al. (\cite{Widrow03}) predicts a mass-to-light ratio for the bulge
($M/L_R=2.7 M_\odot/L_\odot$) that is smaller than the one of the disk
($M/L_R=4.4 M_\odot/L_\odot$), and this despite overestimating the
velocity dispersions along the minor axis. Only model K1, well above
McElroy $\sigma$s, delivers $M/L_R=4.8 M_\odot/L_\odot$ for the bulge.
Widrow et al.  go on discussing why it is so, not taking into account
the possibility that McErloy velocities are in fact systematically too
small.  Of course Widrow et al. model A fails to match our kinematics,
while model K1 slightly overestimates our velocity dispersions.  Fig.
\ref{fig_SPPmodels} shows that for $r>10$ arcsec the analysis of the
Lick indices predicts $M/L_R\approx 4-4.5 M_\odot/L_\odot$, in line
with model K1. As a consequence, the microlensing events estimated by
Widrow et al.  (\cite{Widrow03}), that are based on their model A, are
clearly underestimates.  Similarly, an upward revision of the event
rate estimates of Riffeser et al.  (\cite{Riffeser06}) is needed,
since this paper assumes a stellar mass-to-light ratio for the bulge
of $M/L_R=2.6$. The extremely low $M/L_R$ value (0.8
$M_\odot/L_\odot$) favoured by Chemin et al. (\cite{Chemin09}) based on
a simple mass modeling of the HI velocities is completely ruled out
and points to the inadequacy of underlying modeling assumptions
(circular motions and spherical mass distributions).
%This
%estimate is based on a twice-solar metallicity, uses an IMF different
%from the one used here (with an inner slope of -1.3 for masses
%smaller the $M_{break}=M_\odot$ instead of $0.5 M\_odot$), it includes
%low mass objects down to 0.01 $M_\odot$, but does not account for
%remnants.  Riffeser et al. (\cite{Riffeser08}) include remnants and
%low-mass objects, matching our stellar $M/L_R$ when using
%$M_{break}=0.5 M_\odot$.  In their calculations they have 
%adopted $M/L_R=3.3M\_odot/L_\odot$, obtained for $M_\{break}1 M_\odot$, 
%which is the mass that minimizes the stellar bulge ste.

\begin{figure}%[h!]
\centering
\includegraphics[width=9.5cm]{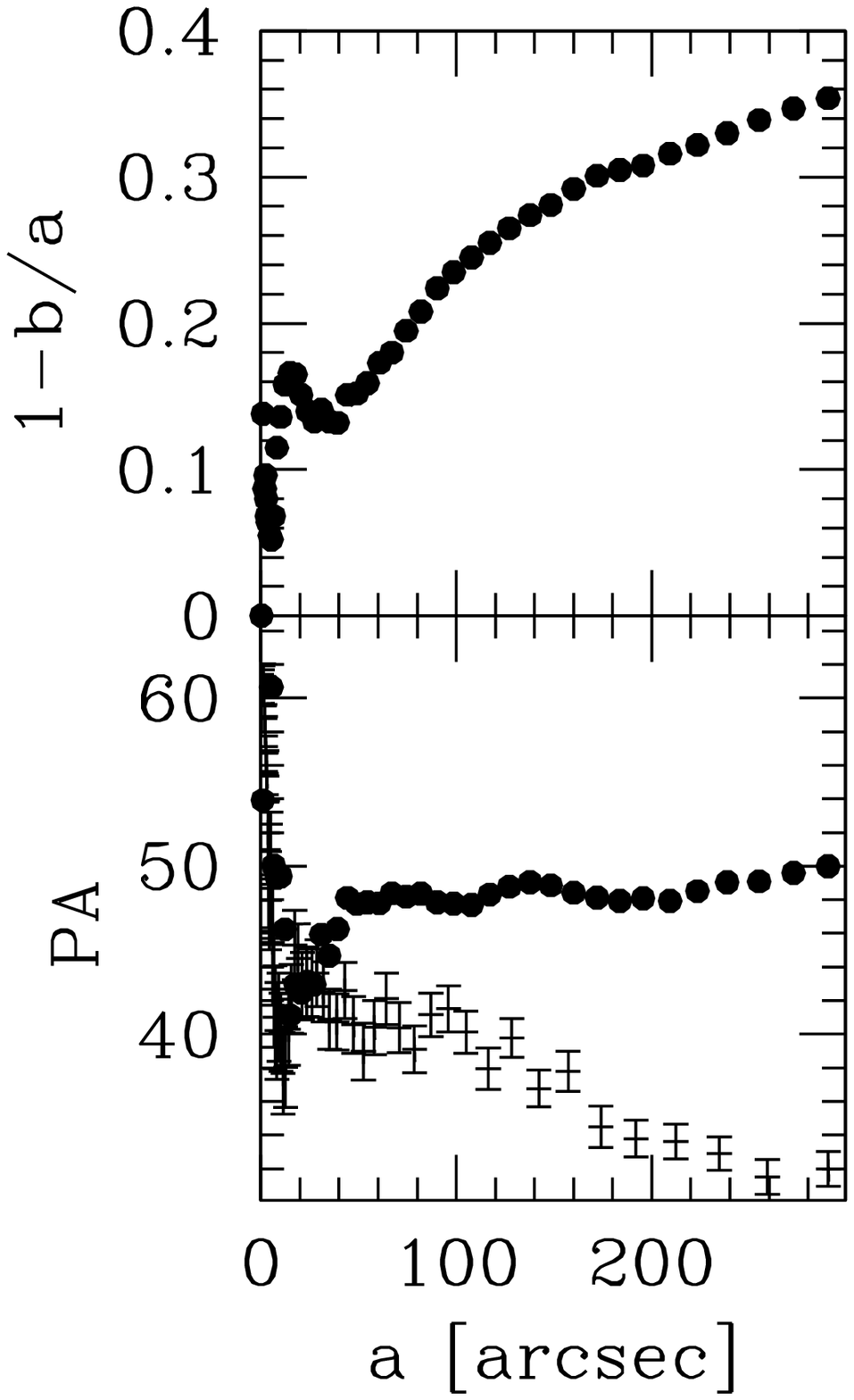}
\caption{Top: the ellipticity profile as a function of the semi-major axis 
distance $a$. Bottom: the position angle of the kinematic major axis 
$\phi_0(a)$ (crosses) and of the isophotes (filled circles) as a function of 
$a$. Both the ellipticity profile and the isophotes PAs are derived from a 
K band 2MASS image of M31.
\label{fig_triaxial}}
\end{figure}

Finally, a word on the triaxiality of the bulge or the possible
presence of a bar. A rotating axisymmetric structure shows equal
velocity curves along slits inclined at positive or negative angles
with respect to the major or minor axis. In general, this does not
happen if the structure is triaxial or a bar is present (see Fig. 12
of Athanassoula \& Beaton \cite{Athanassoula06}). 
In the case of the bulge of M31 the maximal rotation velocity
achieved on the slits M60 and M30 are larger than the ones reached
on slits P60 and P30 respectively, or, in other words, the kinematic
minor axis of the M31 bulge does not coincide with its isophotal
minor axis. We quantify this effect by measuring how the position
angle of the kinematic major axis varies as a function of semi-major
distance. At each semi-major distance $a$ from the center we find
the parameters $V_{max}(a)$ and $\phi_0(a)$ that minimize the
$\chi(a)^2$ function:
\begin{equation}
\chi(a)^2_\phi=\sum_{i=1}^{12}\left[\frac{V_i(a)-V_{max}(a)\cos(\phi_i-\phi_0(a))}
                                  {\delta V_i(a)}\right]^2,
\label{eq_chiphi}
\end{equation} 
where the twelve angles $\phi_i$ correspond to the ones of the slits
shown in Fig. \ref{fig_slits} and listed in Table \ref{tab_log}, and
$V_i(a)$ are the stellar velocities relative to the center shown in
Fig. \ref{fig_kin}, interpolated at the distance proper to the
isophote ellipse determined below (see top panel of Fig.
\ref{fig_triaxial}).  In Eq. \ref{eq_chiphi} we exclude the velocities
measured along the minor axis at distances $r>100$ arcsec, where the
slits are shifted with respect from the center.  A perfectly
axisymmetric system has $\phi_0=PA_{MJ}$, where $PA_{MJ}$ is the
position angle of the isophotal major axis. In general, the function
$V_{max}\cos(\phi_i-\phi_0)$ is a fair representation of the measured
data, with rms deviations of 8 km/s.  Fig.  \ref{fig_triaxial},
bottom, shows the resulting profile $\phi_0(a)$, together with
$PA_{MJ}(a)$ as derived from a 2MASS K image of M31 using the ellipse
fitting package of Bender and M\"ollenhoff (\cite{BM87}). The
corresponding ellipticity profile is shown at the top. The isophotal PA and
ellipticity profiles agree with Beaton et al.
(\cite{Beaton07}). Both the isophotal and the kinematic PAs move with
distance from the center.  But, while the isophotal PA oscillates
around the value of $48^\circ$ (the major axis PA of the M31 bulge,
see Table \ref{tab_log}), the kinematic PA drifts away from it,
approaching $32^\circ$, 6 degrees smaller than the PA of the major
axis of the disk. This suggests that the observed drift is due only
partially to the increasing importance of the stellar disk rotation
and might point to an intrinsic triaxiality of the bulge and/or the presence 
of a bar.

However, only a proper dynamical model that takes into account the
disk contribution, on top of the triaxiality of M31's bulge and its
possible figure rotation, and fits not only the light distribution,
mean velocities and velocity dispersion fields, but also the higher
moments of the LOSVDs, can address these issues quantitatively. We
postpone this discussion to Morganti et al.  (\cite{Morganti09}).

\subsection{The sources of ionizing radiation}
\label{sec_ionize}

Sarzi et al. (\cite{Sarzi09}) discuss in detail how to exploit the
[OIII]/H$\beta$ vs. [NI]/H$\beta$ diagram to constrain the properties
of the ionizing sources for the gas in elliptical and lenticular
galaxies. Their Fig. 1 shows that the diagram neatly separates the
regions where dusty AGNs, shocks and startburst activity are
responsible for the production of emission lines, in analogy with the
standard [OI]/H$\alpha$ vs. [OIII]/H$\beta$ diagram of Veilleux and
Osterbrook (\cite{VO87}).

As discussed by Sarzi et al. (\cite{Sarzi09}), the constancy of the
H$\beta$ EW with radius indicates that the H$\beta$ emission flux
follows the stellar light distribution. To first approximation, this
should apply also to the case discussed here. Values of $\log$
[OIII]/H$\beta\approx 0.5$ with low [NI]/H$\beta$ ratios occupy the
region where shocks are responsible for the gas excitation (see Fig.  1
of Sarzi et al. \cite{Sarzi09}). In contrast, [OIII]/H$\beta$ ratios
as high as 10, like the ones measured in the inner  arcsecs of M31,
fall in the region where photoionization by a central AGN is the
working mechanism.

\section{Discussion and Conclusions}
\label{sec_conc}

We have presented new optical spectroscopic observations of the bulge
of M31. We have measured the stellar and gas kinematics, emission line
strength ratios and absorption Lick indices profiles along 6 position
angles out to distances of 5 arcmin from the center. Along the major
axis we probed regions out to 8 arcmin. We show that the old
kinematics of McElroy (\cite{McElroy83}) provides too small velocity
dispersions (by up to 30\%), therefore biasing the dynamical modeling
towards too small bulge stellar masses (by a factor 2).  Moreover, the
new higher averaged velocity dispersion predicts a mass for the
central supermassive black hole of M31 that is only a factor 2 below
what measured. The new velocity dispersion profiles are now in better
agreement with axisymmetric dynamical models with large bulge
mass-to-light ratio (Widrow et al. \cite{Widrow03}), that now match
the values derived from stellar population models ($M/L_R\approx 4-4.5
M_\odot/L_\odot$, see below). This implies an upward revision of the
predicted self-lensing microlensing event rate of Widrow et al.
(\cite{Widrow03}), and Riffeser et al.  (\cite{Riffeser06}), that are
based on lower stellar mass-to-light ratios.

The inner ($r\le 100$ arcsec) bulge is slowly
rotating, with a $V/\sigma\approx 0.2$. At distances from the center
larger than $\approx 100$ arcsec the measured kinematics becomes
increasingly influenced by the rapidly rotating stellar disk.
Therefore, the observed variation of the kinematic position angle is
suggestive of bulge triaxiality, but needs a proper dynamical modeling
of both disk and bulge components to be quantified. The measured gas
kinematics confirms the well studied large scale disk rotation.
However, a more complex structure, with gas minor axis counter-rotation,
is detected in the inner bulge. This might be evidence for a (recent)
minor merger, possibly connected to the younger stellar population
detected in the inner  arcsecs of the galaxy discussed below.

The analysis of eight Lick index profiles shows that the bulge of M31
is old, of solar-metallicity and a factor 2 overabundant in
$\alpha$-elements, in agreement with studies of its resolved stellar
component (Stephens et al. \cite{Stephens03}). The line indices and
stellar population parameters appear approximately constant on
circles, i.e. their isocontours are rounder than the galaxy isophotes,
as seen in many ellipticals and bulges (Kuntschner et al.
\cite{Kuntschner06}; Falc\'on-Barroso et al.  \cite{Falcon06}).
Together with the derived old ages, this confirms that the stellar
disk out to 5 arcmin from the center is either old (similar to what
found for other spiral galaxies, Peletier and Balcells
\cite{Peletier96}) or not sufficiently probed by the spectral features
considered here. However, we do detect smaller ages ($\le 8$ Gyr) along
the major axis at distances $\ge 6$ arcmin. The u-g SLOAN colors
predicted from our stellar population analysis match reasonably well
with the observed ones. The redder colors g-r and r-i are
systematically offset, a well known problem of the flux calibration of
current simple stellar population models (Maraston et al.
\cite{Maraston09}). The derived mass-to-light ratios (in the Johnson R
band and with a Kroupa IMF we get $M/L\approx 4-4.5$) agree with the
dynamical estimates (see above). They drop to $\le 3 M_\odot/L_\odot$
along the major axis at distances $\ge 6$ arcmin, where the disk light
starts to dominate.

In the inner arcsecs the situation changes and a population with a
light-weighted younger age ($\approx 8$ Gyr inside the seeing disk or
2 arcsec, with values as low as 4 Gyr) and metal richer ($\approx 3$
times solar) appears.  This agrees with the findings of Davidge
(\cite{Davidge97}). In addition, the emission line EW ratios
[OIII]/H$\beta$ increase in this region. Their values are compatible
with being excited by shocks in the main body of the bulge, but near
the center they increase to levels pointing to the presence of an
AGN-like photo-ionizing source. Combined with the detection of
counterrotating gas along the minor axis of the galaxy (see above),
this suggests that a gas-rich minor merger probably happened some 100
Myr ago, that triggered an episode of star formation and possibly
boosted the nuclear activity of the central supermassive black hole of
M31.  We estimate how broad a range of star burst ages and masses can
be to give the measured mean value of 8 Gyr in the inner 2 arcsec,
when superimposed to the old bulge stars background. To this purpose
we compute composite spectra of an old (12.6 Gyr) plus a young (from
100 Myr to 4 Gyr) simple stellar population, using the Vazdekis
(\cite{Vazdekis99}) library, and measure their SSP ages through the
analysis of the Lick indices observed here. We find that global ages
smaller than 8 Gyr are found when considering a young component
younger than $\approx 600$ Myr and a mass fraction lower than 10\%.
Higher mass fractions are possible for older ages. From Kormendy and
Bender (\cite{Kormendy99}), inside an aperture of 2 arcsec diameter we
measure a V mag of 12.5, or $5\times 10^6 L_\odot$. In this region we
estimate $M/L_V\approx 4 M_\odot/L_\odot$ and therefore an enclosed
mass of $2\times 10^7 M_\odot$. As a consequence, $\approx 10^6
M_\odot$ of some 100 Myr old stars would be needed. Note that in
the inner nucleus of M31, at fractions of an arcsec, a disk of 200 Myr
old stars is found (Bender et al.  \cite{Bender05}), with a mass of
$\approx 4200 M_\odot$. Of course the spatial resolution of our
spectra is too low to probe this scale.  Moreover, our result, based
on the H$\beta$ line as an age tracer, depends heavily on the details
of the emission correction and might be affected by the bad column's
interpolation (see Sect. \ref{sec_obs}). Spectra of the higher order
Balmer lines, taken with better seeing, are needed to improve our
conclusions.

In a second paper (Morganti et al. \cite{Morganti09}), a dynamical
model of the data presented here, that takes into account the
contributions of the bulge and disk components, will assess in a
quantitative way the bulge triaxiality issue and will give a proper
estimate of the microlensing event rates due to self-lensing.

\begin{acknowledgements}
  The Hobby-Eberly Telescope (HET) is a joint project of the
  University of Texas at Austin, the Pennsylvania State University,
  Stanford University, Ludwig-Maximilians-Universit\"at M\"unchen, and
  Georg-August-Universit\"at G\"ottingen. The HET is named in honor of its
  principal benefactors, William P.  Hobby and Robert E. Eberly.  The
  Marcario Low Resolution Spectrograph is named for Mike Marcario of
  High Lonesome Optics who fabricated several optics for the
  instrument but died before its completion. The LRS is a joint
  project of the Hobby-Eberly Telescope partnership and the Instituto
  de Astronom\'ia de la Universidad Nacional Aut\'onoma de M\'exico.  The
  grism E2 used for these observations has been bought through the DFG
  grant BE1091/9-1.
\end{acknowledgements}

\end{document}